\documentclass[bibyear]{aa} 
\usepackage{graphicx}
\usepackage{txfonts}
\usepackage{epstopdf}
\usepackage{caption}
\usepackage{subcaption}
\usepackage{color, colortbl}
\usepackage{bm}
\usepackage{array}
\usepackage{hyperref}
\usepackage{siunitx}
\usepackage{xcolor}
\usepackage{natbib}

\newcommand{\CII}{[C{\scriptsize\,II}]}

\newcommand{\OI}{[O{\scriptsize\,I}]}
\newcommand{\HII}{H{\scriptsize\,II}}

\definecolor{Magenta}{rgb}{1,0,1}
\definecolor{LightGray}{gray}{0.9}

\begin{document} 

\title{The Diamond Ring in Cygnus X: Advanced stage of an expanding bubble of ionised carbon}

\author{Simon M. Dannhauer \inst{1}\and
Sebastian Vider \inst{1} \and
Nicola Schneider \inst{1} \and
Robert Simon \inst{1} \and
Fernando Comeron \inst{2} \and 
Eduard Keilmann \inst{1} \and
Stefanie Walch \inst{1} \and
Lars Bonne \inst{3} \and 
Slawa Kabanovic \inst{1} \and
Volker Ossenkopf-Okada\inst{1} \and
Daniel Seifried \inst{1} \and
Timea Csengeri \inst{4} \and
Amanda Djupvik \inst{5,6} \and 
Yan Gong \inst{7,8} \and
Andreas Brunthaler \inst{8} \and
Michael Rugel \inst{9} \and 
Dominik A. Riechers \inst{1} \and
Sylvain Bontemps \inst{4} \and 
Netty Honingh \inst{1} \and 
Urs U. Graf  \inst{1} \and 
A.G.G.M. Tielens \inst{10,11}
}
\institute{I. Physikalisches Institut, Universität zu Köln, Z\"ulpicher Str. 77, 50937 K\"oln, Germany\\
\email{dannhauer@ph1.uni-koeln.de}
\and 
European Southern Observatory, Karl-Schwarzschild-Straße 2, D-85748 Garching, Germany 
\and
SOFIA Science Center, NASA Ames Research Center, Moffett Field, CA 94 045, USA
\and
Laboratoire d’Astrophysique de Bordeaux, Universit\'e de Bordeaux, CNRS, B18N, 33615 Pessac, France
\and
Nordic Optical Telescope, Rambla José Ana Fernández Pérez 7, ES-38711 Breña Baja, Spain 
\and
Department of Physics and Astronomy, Aarhus University, Munkegade 120, DK-8000 Aarhus C, Denmark
\and
Purple Mountain Observatory, and Key Laboratory of Radio Astronomy, Chinese Academy of Sciences, 10 Yuanhua Road, Nanjing 210023, PR China
\and
Max-Planck Institut f\"ur Radioastronomie, Auf dem H\"ugel 69, 53121 Bonn, Germany 
\and
 National Radio Astronomy Observatory, PO Box O, 1003 Lopezville Road, Socorro, NM 87801, US
\and
Department of Astronomy, University of Maryland, College Park, MD 20742, USA
\and
Leiden Observatory, Leiden University, PO Box 9513, 2300 RA Leiden, The Netherlands 
}
        
\date{draft of \today}
\titlerunning{The Diamond Ring in Cygnus X}  
\authorrunning{S. Dannhauer}  

\abstract{
The `Diamond Ring' within Cygnus~X, south-west of the DR21 ridge, 
stands out as a prominent, nearly circular structure in infrared (IR) and far-infrared (FIR) emission, spanning approximately 6~pc in diameter. It is enclosed by clumpy molecular clouds seen in CO lines and contains an \HII\ region, visible in cm emission. It resembles a classical \HII\ region associated with an expanding bubble seen routinely in the 158 $\mu$m line of ionised carbon (\CII).
However, our recent observations utilising the Stratospheric Observatory for Far-Infrared Astronomy (SOFIA) under the FEEDBACK program for the spectrally resolved \CII\ line have revealed 
a slightly tilted ring with a mass of $\sim$10$^3$ M$_\odot$, advancing at a velocity of $\sim$1.3 km s$^{-1}$. 
The bulk emission of the gas has a line-of-sight (LOS) velocity around $-$2 km s$^{-1}$. 
The \CII\ data revealed that the `Diamond' of the Diamond Ring is an unrelated, dense gas clump at a LOS velocity of $\sim$7 km s$^{-1}$. 
The driving source, which is also responsible for powering the associated \HII\ region, is a B0.5e star, classified by our IR spectroscopy.  
This observation marks the first instance where we observe only a slowly expanding ring of \CII\  emission and not an expanding 3D shell. 
We suggest that the \HII\ region (along with its associated \CII\ bubble), initially formed by a massive star, expanded outwards from a flat slab of molecular gas nearly in the plane of the sky. 
Presently, the \CII\ ring is confined by the swept-up gas of the slab, while the parts of the shell moving in the directions perpendicular to the shell along the LOS have dissipated, resulting in a notable decrease in the expansion of the remaining ring. This scenario is supported by dedicated simulations that trace the evolution of the \CII\ bubble. 
Our observations support the scenario of \HII\ region evolution in `flat' molecular clouds, reported earlier in the literature. In this geometry, we propose that the Diamond Ring represents the terminal phase of an expanding \CII\ bubble driven by stellar winds and thermal pressure.}    
                
\maketitle

\section{Introduction} \label{sec:intro} 

\begin{figure}[htbp]
\begin{center} 
\includegraphics [width=8cm, angle={0}]{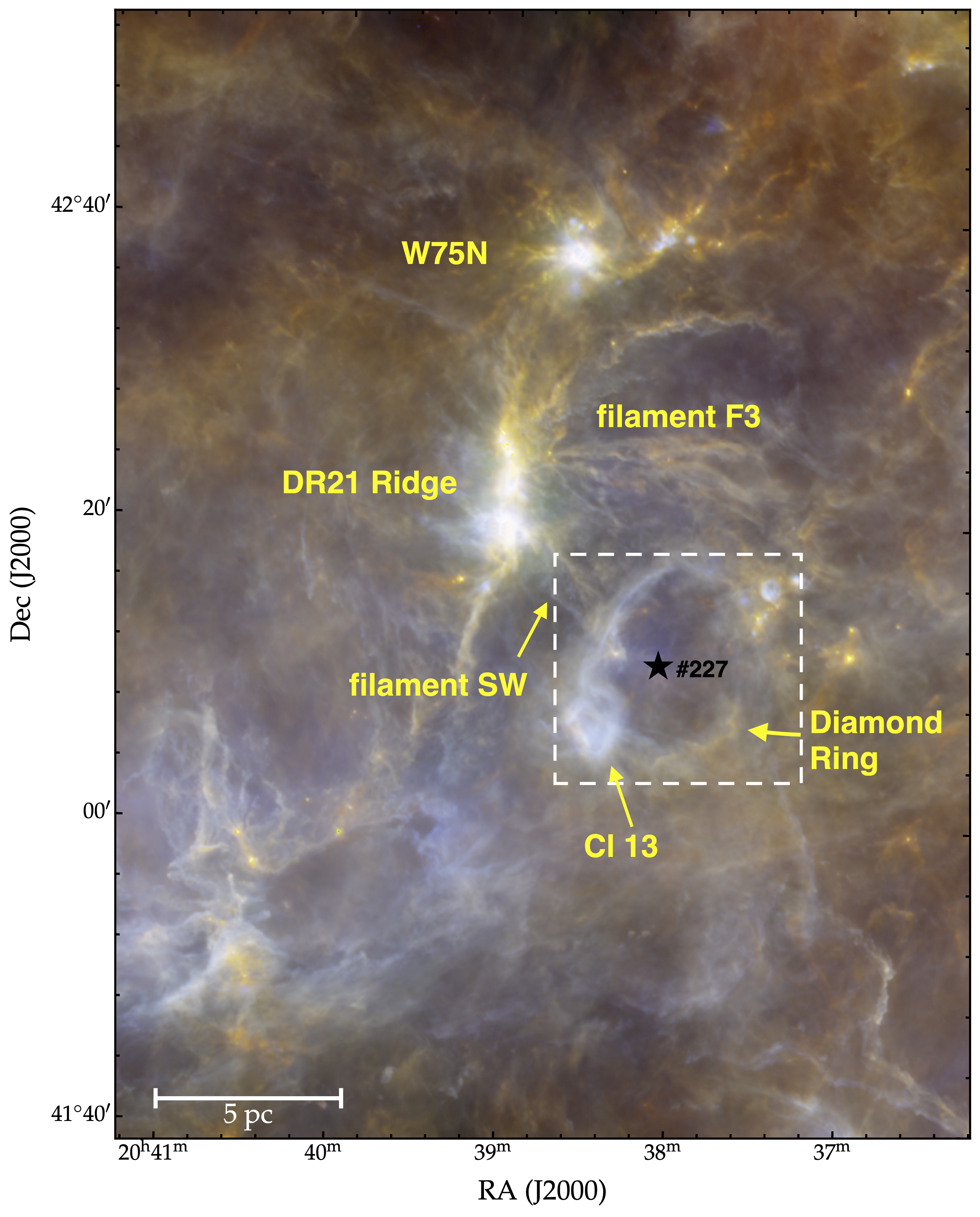}
\caption{Overview of the northern Cygnus X region. The three colour image displays {\sl Herschel} 70 $\mu$m (blue), 160 $\mu$m (green), and 250 $\mu$m (red) emission \citep{Schneider2016a}.  The 70 and 160 $\mu$m emissions indicate the distribution of hot dust while the 250 $\mu$m trace the cooler dust mostly associated with molecular gas.  
The main objects are labelled in yellow, the filament notation stems from \citet{Hennemann2012}. 
The B star \#227 is indicated with a black star symbol. The `Diamond' of the Diamond Ring is Cl 13  \citep{LeDuigou2002}. The \CII\ mapping covers the DR21 ridge and the Diamond Ring \citep{Schneider2023,Bonne2023}. Here, we focus  on the area outlined by a white square.}
\label{fig:herschel}
\end{center} 
\end{figure}

\begin{figure*}[htbp]
\begin{center} 
\includegraphics [width=8.6cm, angle={0}]{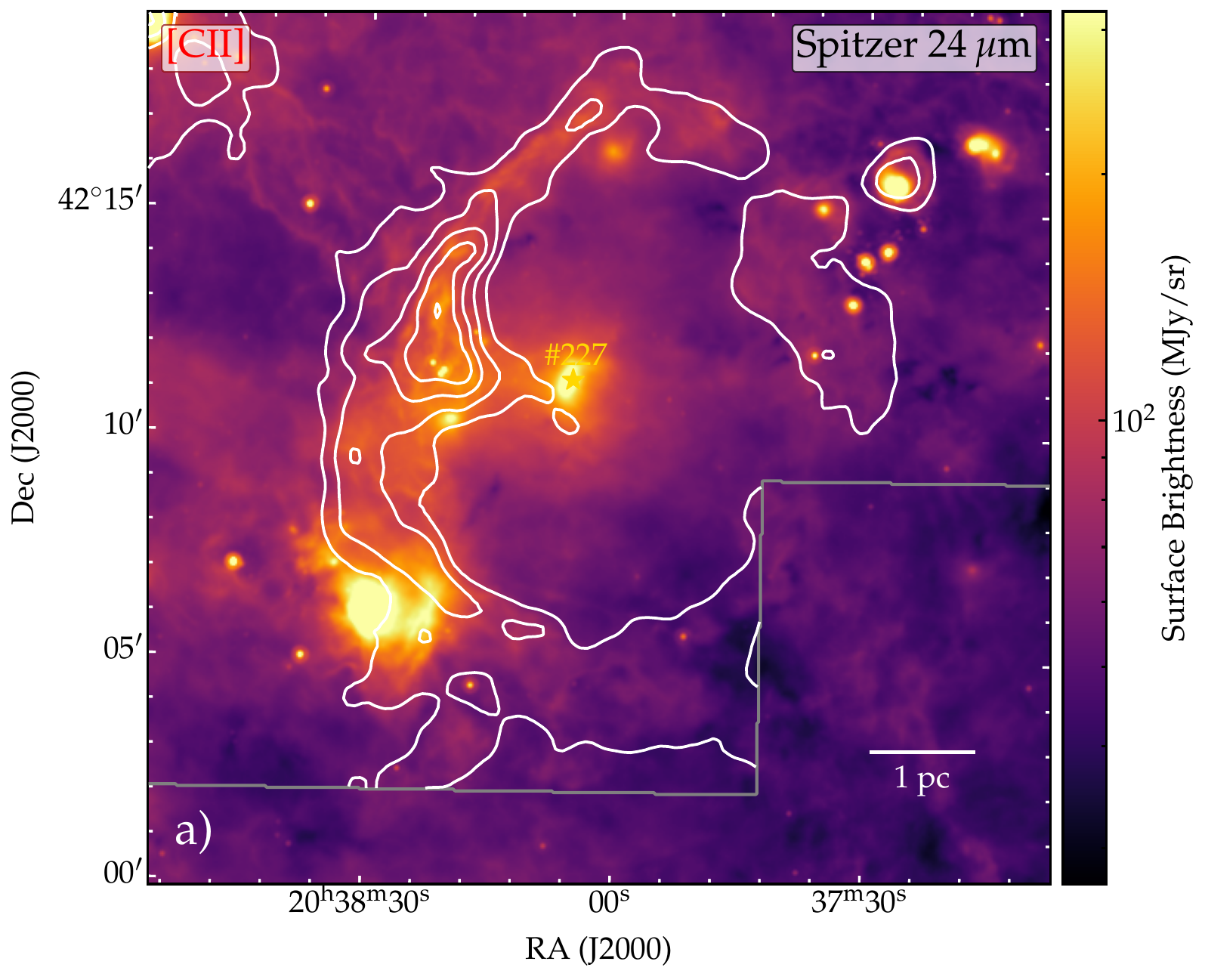}
\includegraphics [width=9cm, angle={0}]{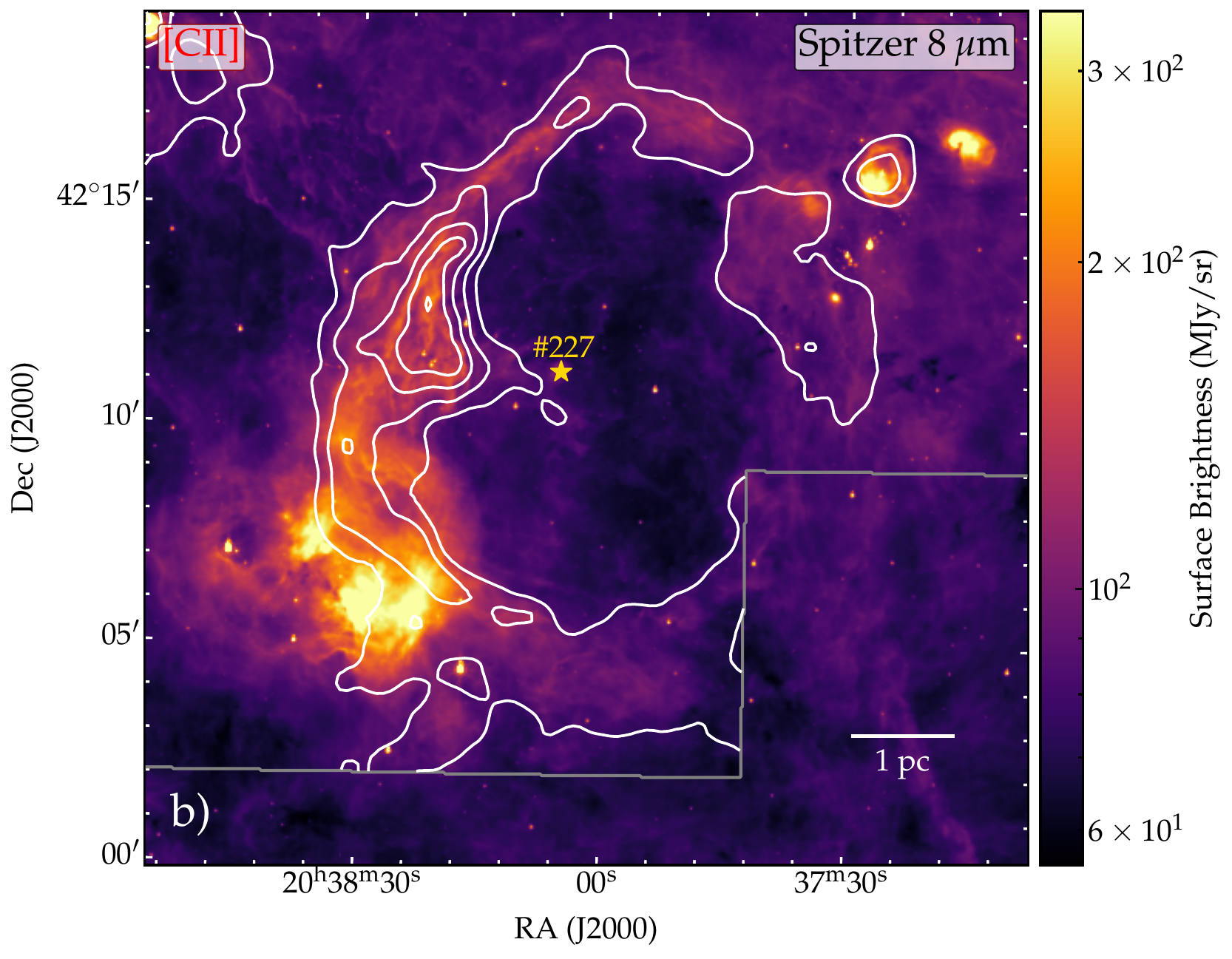}
\includegraphics [width=8.6cm, angle={0}]{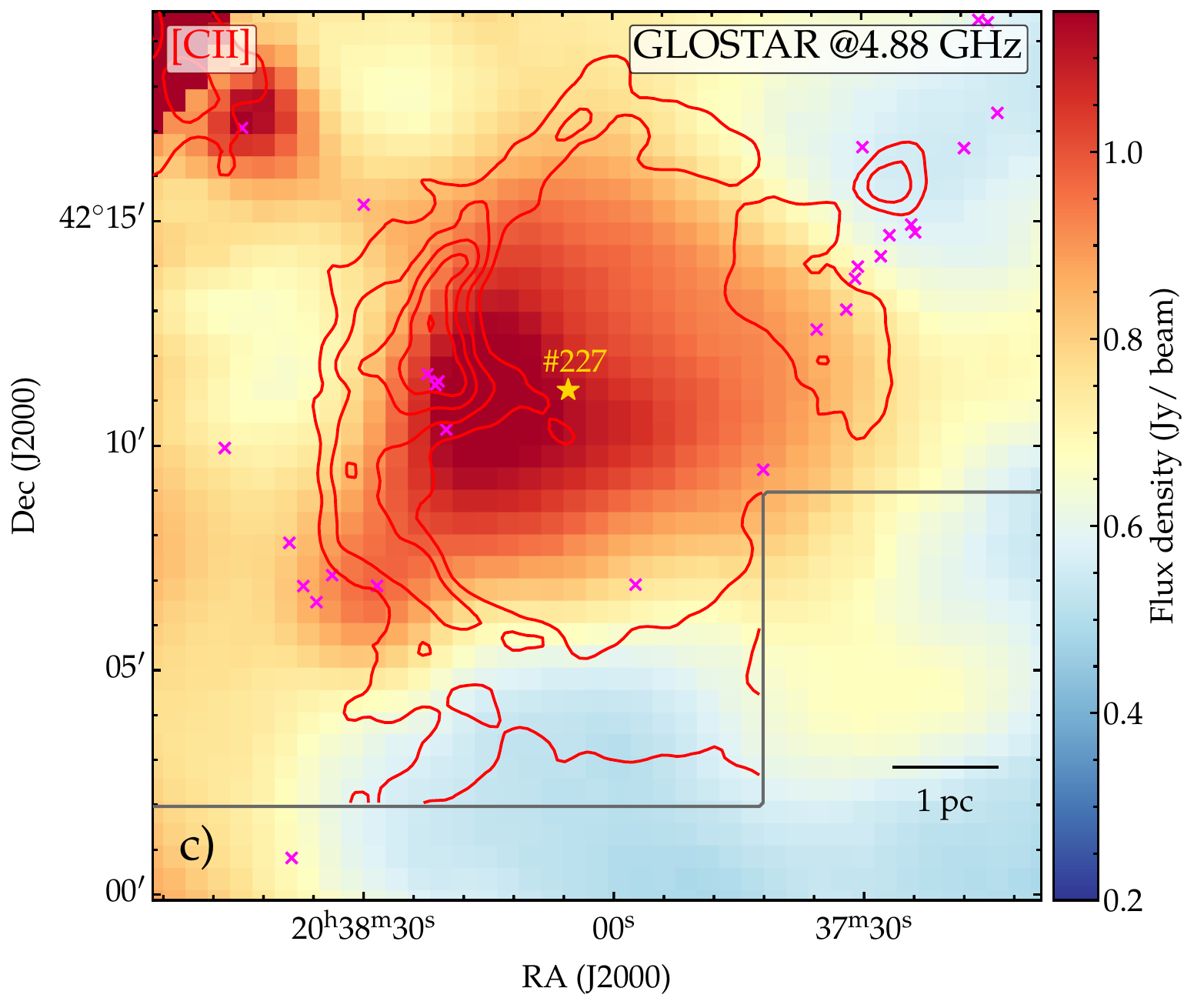}
\includegraphics [width=9cm, angle={0}]{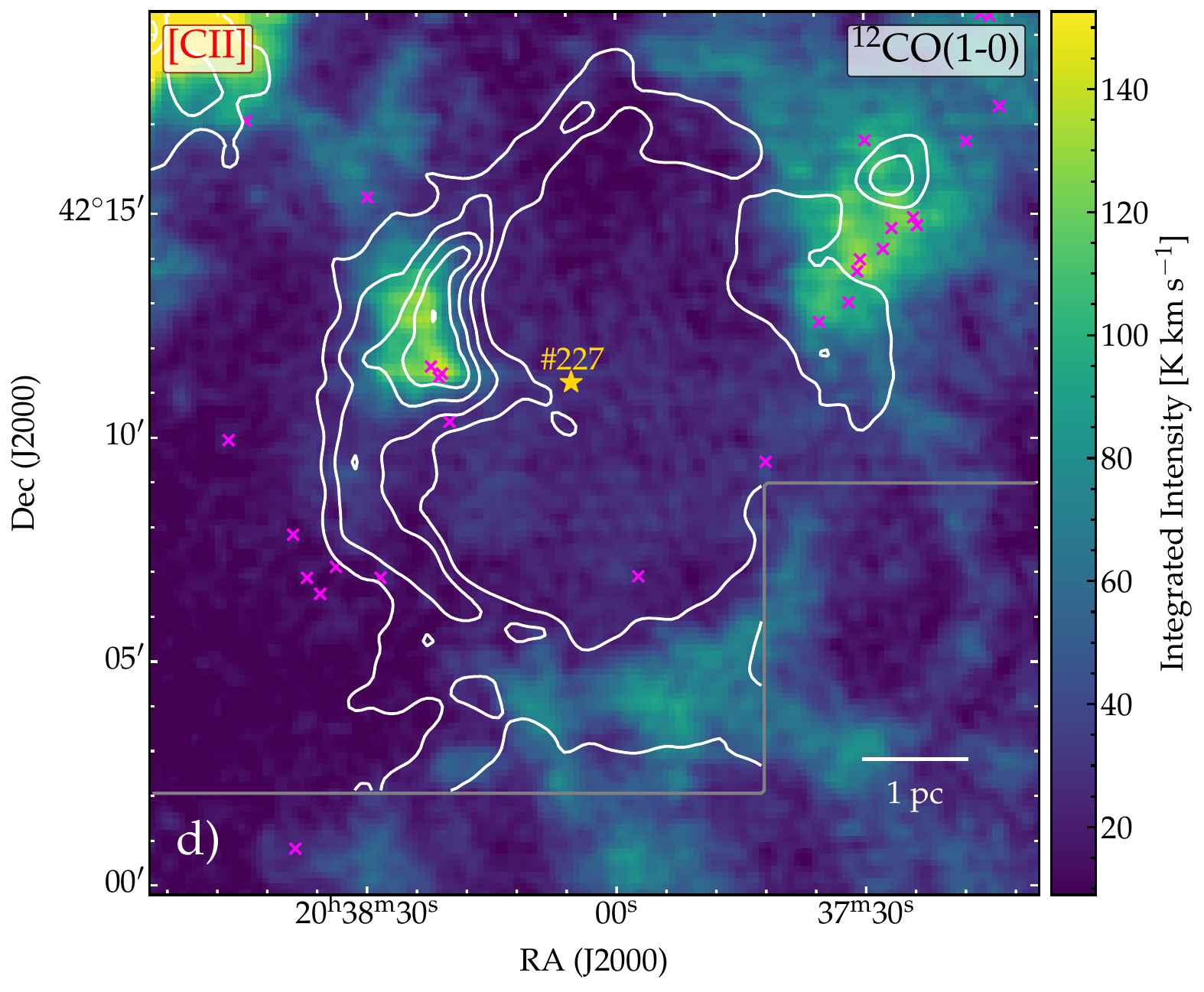}
\vspace{-0.2cm}
\caption{Multi-wavelength overview of the Diamond Ring. {\sl Spitzer/MIPS} 24 $\mu$m (a) and 8 $\mu$m (b) and 4.88 GHz GLOSTAR (c) Effelsberg-only (145$''$ beam) emission of the \HII\ region associated with the Diamond Ring. Panel d) shows the $^{12}$CO 1$\to$0 emission in the velocity range $-$8 to 3 km s$^{-1}$. The overlaid line integrated ($-$8 to 3 km s$^{-1}$) \CII\ contour levels (40 to 120 K km s$^{-1}$ in steps of 20 K km s$^{-1}$) in all panels correspond only to the emission of the Diamond Ring (and not cluster Cl~13). The position of the ionising star \#227 is indicated with a yellow star, the position of the protostars of \cite{Kryukova2014} are marked by magenta crosses in (c) and (d). They are amply visible in the 24 $\mu$m image. The south-western corner of the DR was not observed in \CII\, as indicated by the gray line.}
\label{fig:hii-region}
\end{center} 
\end{figure*}

The radiation and stellar winds emitted by OB stars inject substantial radiative, thermal, and mechanical energy into the interstellar medium (ISM). 
In recent years, a highly effective method for quantifying stellar feedback has been to utilise the fine-structure line of ionised carbon (\CII) at 158 $\mu$m because it probes photo-dissociated regions (PDRs), where far-ultraviolet (FUV) photons predominantly determine the physical and chemical properties of the gas \citep{Hollenbach1999}. 
Notably, observations conducted with the Stratospheric Observatory for Infrared Astronomy (SOFIA) have enabled velocity-resolved mapping of \CII\ around \HII\ regions, revealing the presence of radially expanding \CII\ shells \citep{Pabst2019,Luisi2021,Tiwari2021,Beuther2022,Bonne2022}.
They are most visible as high-velocity red- or blueshifted emission in \CII\ spectra and display an ellipse in position-velocity (PV) cuts. The maximum expansion velocities along the line-of-sight (LOS) in the central part of the bubble typically surpass the sound speed of approximately 10 km s$^{-1}$, characteristic of photo-ionised gas. This led to the conclusion that the \CII\ bubbles are predominantly driven by the stellar winds of the exciting OB star(s), as depicted in analytical models \citep{Weaver1977}. Lower expansion velocities of evolved \CII\ shells have also been observed in Orion \citep{Pabst2020}. 
However, only recently, \citet{Keilmann2025} detected a very young \CII\ bubble characterised by a small compact \HII\ region and an extended, slowly expanding ($\sim$2.6~km~s$^{-1}$)  \CII\ shell. 

We report the first detection of a distinct ring\footnote{
We observe a ring as a projected structure on the sky. As we will demonstrate later, the parental molecular cloud likely exhibits a slab-like morphology. Consequently, in three-dimensional space, the overall geometry is best described as a toroidal structure.} of \CII\ emission within the Cygnus~X region that is devoid of any spherical shell-like configuration, but displaying, in fact, a radial expansion in the plane of the sky at $\sim$1.3 ${\rm km\,s^{-1}}$. This feature corresponds to the  `Diamond Ring' \citep{Marston2004}, situated south-west of the DR21 ridge \citep{Schneider2010}, cf. the {\sl Herschel} map displayed in Fig.~\ref{fig:herschel}.  The whole area forms a complex network of clouds and filaments, hypothesised to result from the interaction of predominantly atomic flows \citep{Schneider2023}. All western filaments, in particular, filament SW (notation from \citet{Hennemann2012}), reveal a connection to the ring. 
The Diamond Ring (or at least parts of it) is also conspicuous in the near- and mid-infrared (NIR-MIR) \citep{Marston2004,Beerer2010,Kryukova2014}. In Fig.~\ref{fig:hii-region} a and b, we  map the  {\sl Spitzer} 24 and 8~$\mu$m data.  It is also discernible in CO transitions (see Fig.~\ref{fig:hii-region} d) and the \CII\ line \citep{Schneider2023}. \\ 
The nearby Cyg~OB2 association with 169 O stars \citep{Wright2015} illuminates the entire complex. The Diamond Ring itself is associated with an \HII\ region that fills the cavity \citep{Lockman1989,Kuchar1997} with its exciting source most plausibly being star \#227 from \citet{Comeron2008}. A NIR spectroscopy study for of this star is also presented in the present paper. 
Figure \ref{fig:hii-region} c shows a map taken at 4.88 GHz (6 cm), where it becomes obvious that the ionised gas fills the Diamond Ring, with stronger emission in the eastern part. This could be due to a density gradient within the \HII\ region and radiation leakage from the embedded cluster located in the eastern clump within the Diamond ring. The protostellar sources within this clump \citep{Kryukova2014} are well visible in the 24 $\mu$m image in Fig. \ref{fig:hii-region} a). \\ 
The Diamond Ring has the same velocity as the molecular clouds associated with the DR21 ridge at $\sim -2\,{\rm km\,s^{-1}}$ in the Cygnus X North region, so we adopted a distance of $1.5\,{\rm kpc}$ \citep{Rygl2012}. Here, we demonstrate that the `Diamond' component of the Diamond Ring in the south-east, containing a cluster of B stars referred to as Cl~13 in \citet{LeDuigou2002} or cluster~16 in \citet{Dutra2001}, is a LOS feature unrelated to the ring at a velocity of around +7 km s$^{-1}$. 
The objective of this paper is to understand how this peculiar \CII\ ring has been formed and how it evolves. To achieve this aim, we use \CII\ data from the SOFIA legacy program FEEDBACK \citep{Schneider2020}, along with archival molecular line, dust continuum and radio data, and new optical observations of star \#227. The SOFIA and optical observations are described in Sect.~\ref{sec:obs}. All data sets are presented and analysed in Sect.~\ref{sec:results}. We also performed a dedicated 3D simulation, informed by observational parameters, where the radiation and stellar wind of a massive star erode a slab of gas (Sect.~\ref{sec:discuss}). In Sect. 4, we also develop a physical scenario for the Diamond Ring and discuss the observed timescales. Section~\ref{sec:summary} summarises the paper.

\section{Observations} \label{sec:obs} 

\subsection{SOFIA} \label{sec:sofia} 

The Diamond Ring was mapped as part of a large-scale mapping of the Cygnus X north region within the SOFIA FEEDBACK legacy program. \footnote{\url{https://feedback.astro.umd.edu} and \url{https://astro.uni-koeln.de/index.php?id=18130}} 
The \CII\ 158 $\mu$m line was observed using the upGREAT receiver \citep{Risacher2018}. The half-power beam width at 158 $\mu$m is 14.$''$1 and the typical main beam efficiency is approximately 0.65. We use data that we spatially smoothed to angular resolutions of 20$''$, on a Nyquist sampled 8$''$ grid, and 15$''$ in a 5$''$ grid. The 20$''$ spectra were resampled to a velocity resolution of $0.3\,{\rm km\,s^{-1}}$ and 1 km s$^{-1}$ and have an average rms baseline noise of 0.8 K and 0.5 K at these velocity resolutions. At 15$''$ angular resolution the average rms baseline is 1.05 K at $0.3\,{\rm km\,s^{-1}}$ velocity resolution. 
For further technical details, we refer to  \citet{Schneider2020,Schneider2023} and \citet{Bonne2023}. 

\subsection{Spectroscopy with the Nordic Optical Telescope} \label{sec:not} 
Spectroscopy of star \#227 was obtained on the night of 27 and 28 June 2024 with the NIR camera and spectrograph NOTCam at the Nordic Optical Telescope (NOT) at La Palma, Spain. Two separate settings were used, covering the H band (1.48-1.78 $\mu$m) and the K band (1.95-2.36 $\mu$m). The slit used had a width of 0$''$6, providing a resolution 
$\lambda/\Delta \lambda$=2500. The star was moved along the slit using an ABBA pattern and in each position, a set of ten exposures of 4~s was obtained in each filter. The exposure time was thus 160~s in each filter, which was sufficient to obtain a signal-to-noise (S/N) ratio of 80 in the H band and 170 in the K band, respectively. A correction of telluric features was achieved by dividing the spectra of star \#227 by that of the nearby B8V star HD 194670 subsequently observed at the same airmass.

\begin{figure*}[htbp]
\begin{center} 
\includegraphics [width=9.0cm, angle={0}]{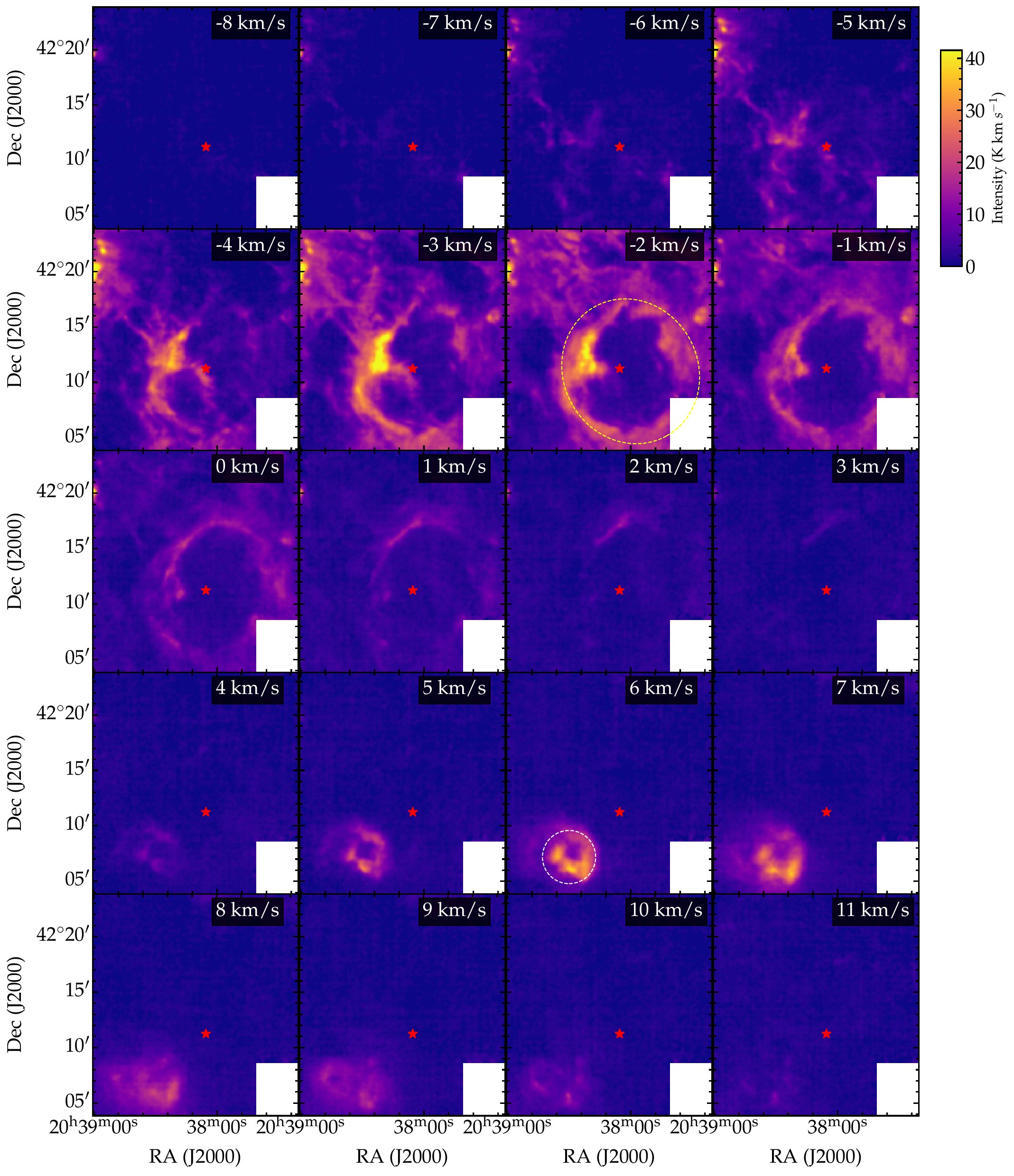}
\includegraphics [width=9.0cm, angle={0}]{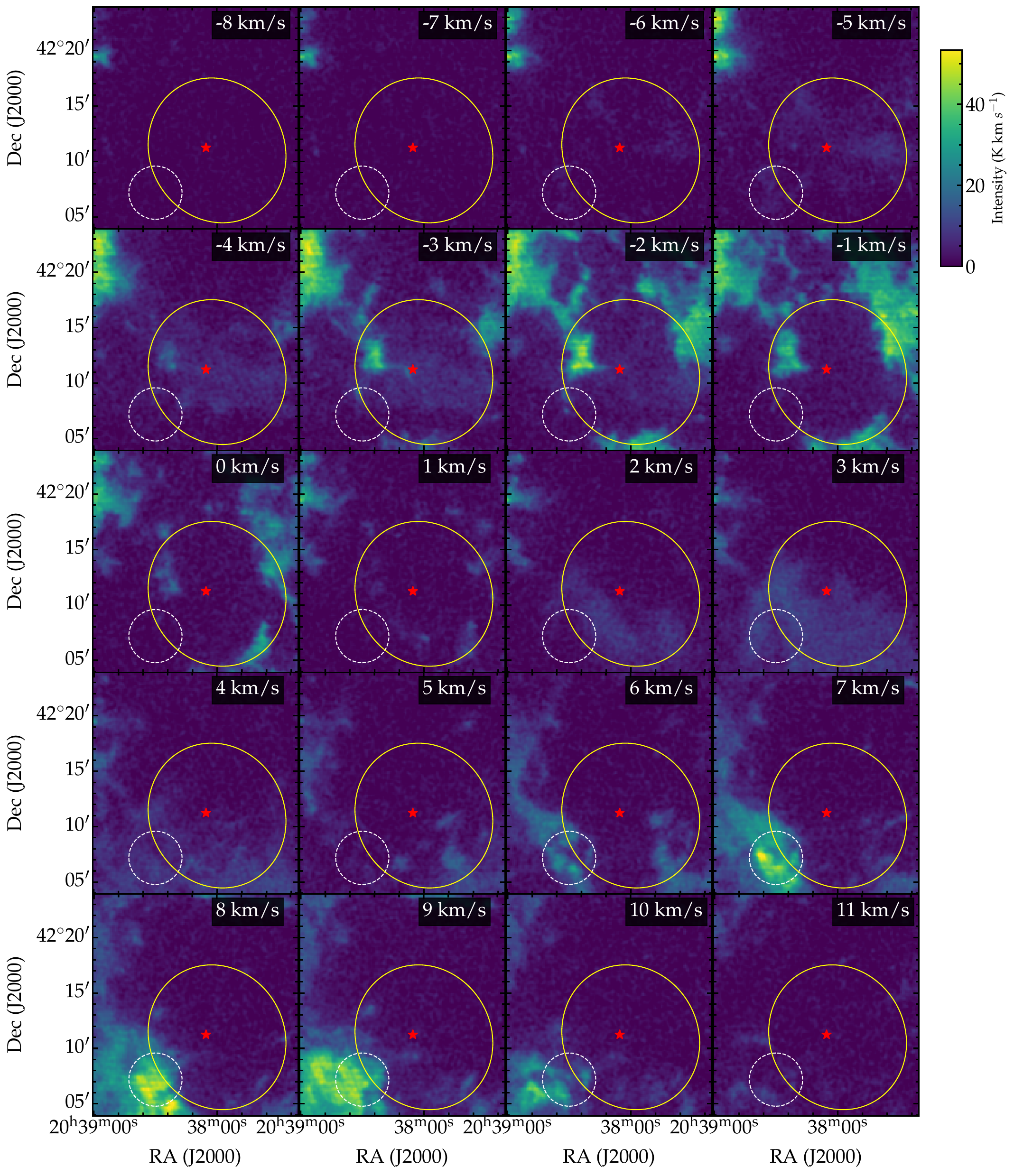}
\vspace{-0.2cm}
\caption{Channel maps of \CII\ (left) and CO 1$\to$0 (right) emission. A yellow ellipse and a white circle indicate the positions of the Diamond Ring and Cl~13, respectively.
The velocity spacing is 1 km s$^{-1}$ at an angular resolution of 20$''$, the position of star \#227 is shown by a red stellar symbol. 
} 
\label{fig:channels}
\end{center} 
\end{figure*}

\begin{figure}[htbp]
\begin{center} 
\includegraphics [width=8cm, angle={0}]{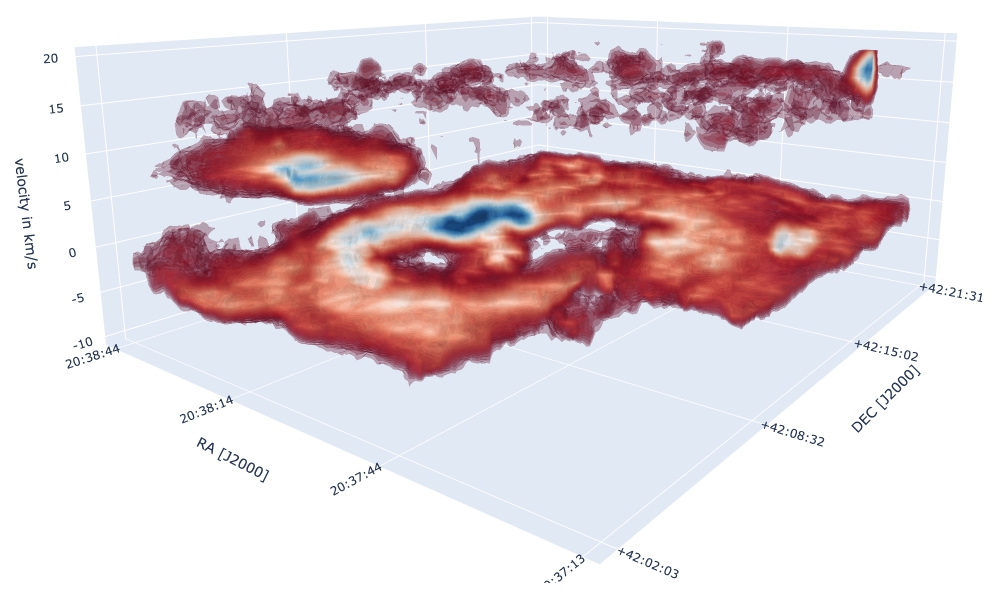}
\vspace{-0.2cm}
\caption{3D position velocity cut of \CII\ emission of the Diamond Ring. The angular resolution is 30$''$, the velocity resolution is 2 km s$^{-1}$, and the colour scale ranges from 5 to 20 K km s$^{-1}$. 
No distinct `bubble' feature is observed; the emission is concentrated around the bulk velocity of $-$2 km s$^{-1}$. The \CII\ emission associated with Cl~13 at 7 km s$^{-1}$ is clearly discernible. Additionally, more diffuse emission is detected at 15 km s$^{-1}$, described as high-velocity gas in \citet{Schneider2023}.
An interactive version is accessible online.
}
\label{fig:Diamond-3D-PV}
\end{center} 
\end{figure}

\begin{figure}[htbp]
\begin{center} 
\includegraphics [width=9cm, angle={0}]{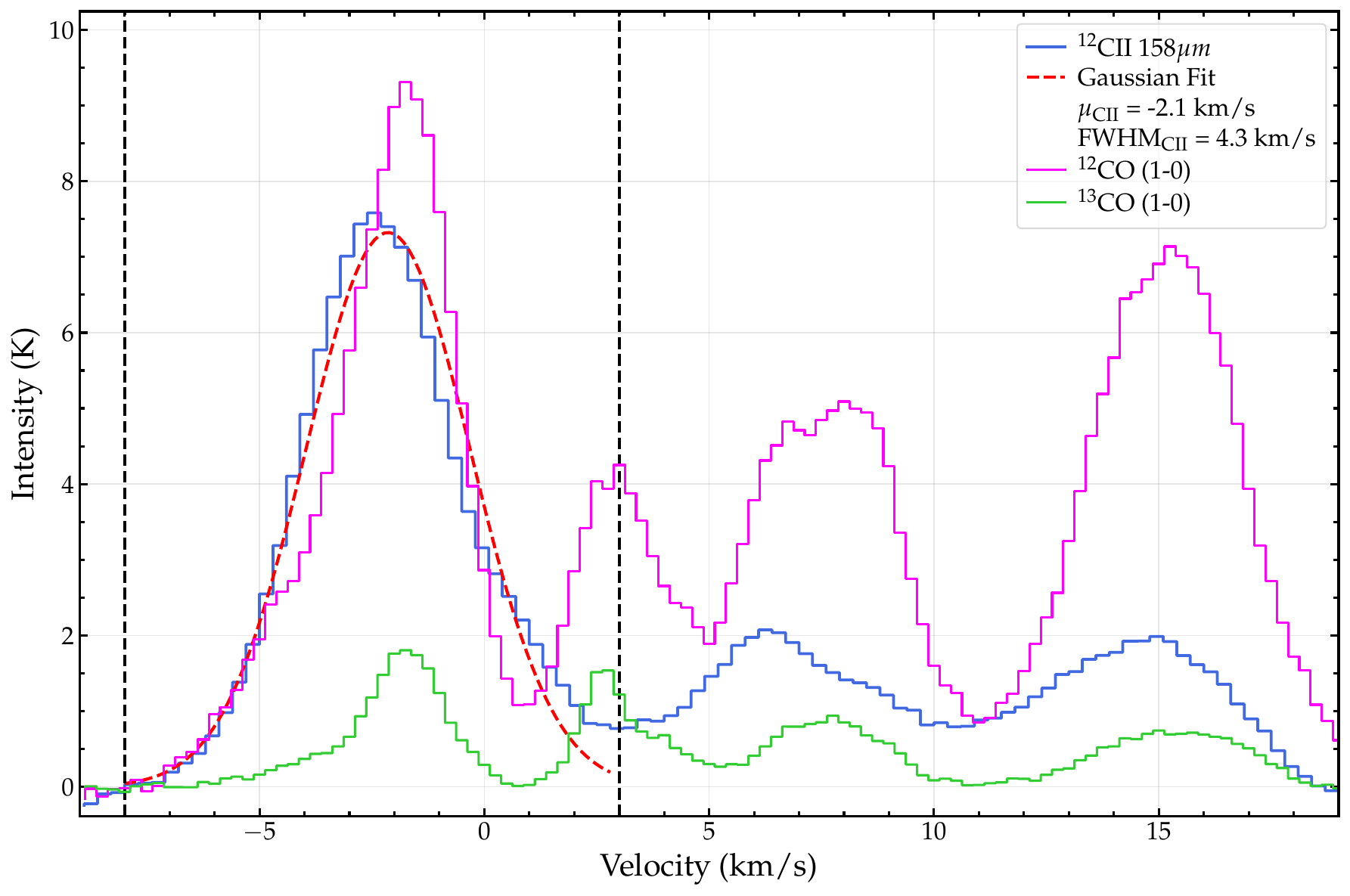}
\caption{Average spectra across the Diamond Ring. The \CII, $^{12}$CO and $^{13}$CO 1$\to$0 line emissions, indicated with different colours, were averaged over a circular area of radius 8.3$^{'}$ cantered on the Diamond Ring.  The velocity range corresponding to the \CII\ ring emission (-8~km s$^{-1}$ to 3 km s$^{-1}$) is marked the by vertical dashed lines (left). The peak emission of the ring is at -2.1~km s$^{-1}$, as determined by a Gaussian fit. The peaks at 7~kms$^{-1}$ and 15~km s$^{-1}$ correspond to Cl13 and the high-velocity gas, respectively \citet{Schneider2023}.
}
\label{fig:spectra}
\end{center} 
\end{figure}

\subsection{Complementary data} \label{sec:lines} 
We used $^{12}$CO and $^{13}$CO 1$\to$0 data from the Nobeyama 45m telescope Cygnus survey, described in detail in \citet{Yamagishi2018} and \citet{Takekoshi2019}. The original angular resolution is 16$''$ and the velocity resolution 0.25 km s$^{-1}$; however, for the purposes of this study, we used data smoothed to 20$''$ resolution in a 8$''$ grid.
In addition, we employed 8 $\mu$m and 24 $\mu$m data at 1.9$''$ and 2.55$''$ angular resolution, respectively, from the {\sl Spitzer} Cygnus X legacy survey \citep{Kryukova2014} and FIR continuum images from {\sl Herschel}, shown in \citet{Schneider2016a}. 

Finally, we also made use of radio continuum data at 4.88 GHz from the GLOSTAR \citep{Brunthaler2021,Gong2023} survey, which combines data taken at the Effelsberg 100m telescope and the VLA. For technical details, we refer to these papers. Here, we use an Effelsberg-only map with a resolution of 145$''$. 

\section{Results and analysis} \label{sec:results}

\subsection{The kinematics of the Diamond Ring} \label{subsec:kine}

\noindent Figures~\ref{fig:hii-region}, ~\ref{fig:channels}, and \ref{fig:Diamond-3D-PV}  and 2D PV  plots in Appendix \ref{appendix-plots} explore the emission distribution and velocity structure of the \CII\ and CO lines in the Diamond Ring. All figures reveal a mostly coherent ring-like structure in \CII\ at the DR21 bulk velocity, spanning from $-$3 km s$^{-1}$ to 1 km s$^{-1}$. 
Figure~\ref{fig:spectra} presents the average spectrum across the Diamond Ring region in the \CII, $^{12}$CO, and $^{13}$CO 1$\to$0 lines. The spectrum reveals a complex velocity structure, with emission originating from previously identified molecular cloud complexes in Cygnus X \citep{Schneider2006, Cao2019, Schneider2023}. We focus on the velocity range from $-8$ to 3 km s$^{-1}$, which corresponds to the Diamond Ring.
We did not detect the $[{}^{13}\mathrm{C}\,\mathrm{II}]$ hyperfine structure components F(1-0) and F(1-1), which have velocity shifts of $-$65.1~km~s$^{-1}$ and +63.2~km~s$^{-1}$, respectively, relative to the bulk emission at $-$2~km~s$^{-1}$. The F(2-1) component has a velocity shift of 11.3 km s$^{-1}$ and, thus, it would be expected to appear around 7 km s$^{-1}$. This makes the line difficult to detect because  of the bulk emission of the Diamond with Cl~13 at that velocity.

The PV cuts and channel maps clearly demonstrate that the `Diamond' of the Diamond Ring,  identified as Cl~13 in \citet{LeDuigou2002}, is detached from the ring in velocity space, with its bulk emission centred at 7 km s$^{-1}$. The peak of the \CII\ emission originates from a clump located east of star \#227, which exhibits a protrusion extending toward this source. The clump contains four protostellar objects with luminosities above 10 L$_\odot$ \citep{Kryukova2014}. 
We note that the CO (1$\to$0) emission (Figs.~\ref{fig:hii-region} d) and \ref{fig:channels}) is very fragmented at all velocities and does not form a coherent ring-structure, as we see in \CII. Moreover, the CO emission appears to wrap around the \CII\ emission and the ring. 
The channel maps (Fig.~\ref{fig:channels}) show that at $-$2 km s$^{-1}$, the SW filament (Fig.~\ref{fig:herschel}) is clearly observed in CO (while more diffuse in \CII), while at 7 km s$^{-1}$, the molecular cloud associated with Cl~13 becomes prominent. \\

\begin{figure*}[htbp]
\begin{center} 
\includegraphics[width=8cm, angle=0]{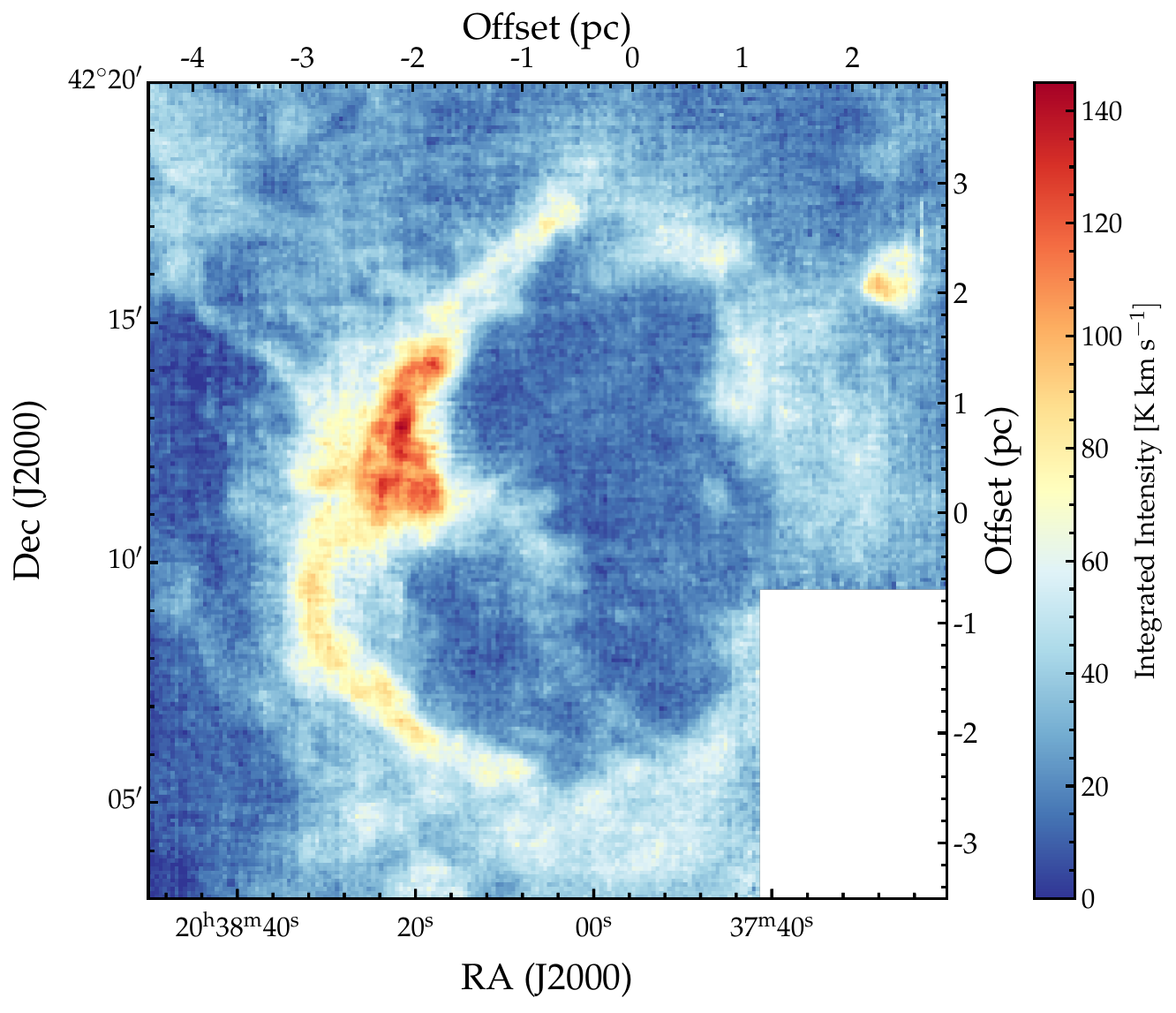}
\includegraphics[width=8cm, angle=0]{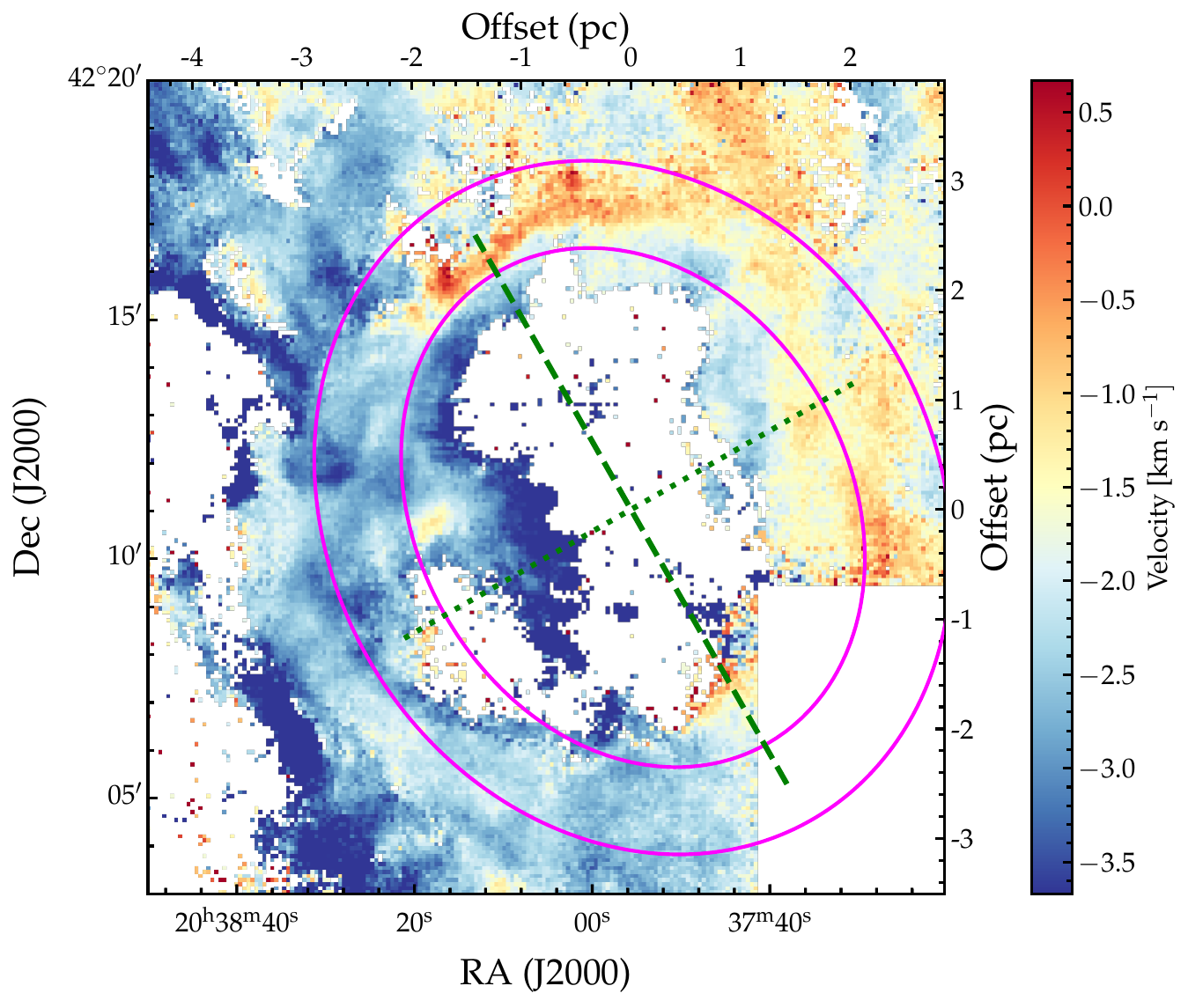}
\vspace{-0.2cm}
\caption{Moment-0 and moment-1 map of the observed \CII\ emission. The moment-0 map is integrated over a velocity range of $-$8 to 3 km s$^{-1}$.
The green dashed lines in the moment-1 map indicate the major and minor axes and the pink lines outlines the 7~K contour of the fitted ellipse to guide the reader's eye. Offsets are given with respect to the centre of the same ellipse. Details of the fitting procedure are given in Appendix~\ref{appendix-ellipse}. The moment-2 map (line width distribution) is given in Fig.~\ref{fig:ellipse_moment2}. } 
\label{fig:ellipse_moment1}
\end{center} 
\end{figure*}

The most important aspect to consider here is the lack of significant blue- or redshifted \CII\ emission in the centre of the ring at high velocities (at least a few km s$^{-1}$ away from the bulk emission), which is typically associated with the 3D expansion of a 
radiation or wind-blown bubble (see references in Sect.\ref{sec:intro}). Our 1$\sigma$ detection limit is 0.52 K km s$^{-1}$ in the channel map of Fig.~\ref{fig:channels} and 1.94 K km s$^{-1}$ for the line-integrated ($-$8 to 3 km s$^{-1}$) \CII\ map. The noise level in the channel map corresponds to a typical \CII\ column density of N$_{\rm CII}$~=~$9.4\cdot10^{15}$ cm$^{-2}$ (following the equations presented later). 
A possible \CII\ shell, however, would only be visible at high red- and blueshifted velocities and would cover a smaller velocity range (not including the bulk emission). 
Typical line widths of expanding \CII\ shells are around 5 km s$^{-1}$ \citep{Luisi2021,Bonne2023b}, leading to N$_{\rm CII}$~=~$2.4\cdot10^{16}$ cm$^{-2}$. We observed slightly blueshifted ($\sim -$2 to $-$3 km s$^{-1}$ LOS velocity) emission in the southeastern region and redshifted ($\sim -$1.5 to 0.5 km s$^{-1}$) emission in the northwestern region, as displayed in Fig.~\ref{fig:ellipse_moment1}. 
An increased  velocity dispersion (Fig.~\ref{fig:ellipse_moment2}) is observed in the northeastern part of the Diamond Ring, where the SW filament intersects the ring. Ram pressure caused by collision with this filament could explain the higher velocity dispersion. However, this will be investigated in a separate study, using recently obtained molecular line data (Dannhauer et al., in prep.). In contrast, the higher velocity dispersion  further south in the ring is likely a consequence  of enhanced turbulence induced by stellar feedback from the internal B-type stars within the clump located east of star \#227. The higher \CII\ intensity of up to 140 K km s$^{-1}$ in this clump (compared to typical values of 60-70 K km s$^{-1}$ in other regions of the Diamond Ring) is most likely also caused by internal stellar feedback.  

We assume that the elliptical appearance of the Diamond Ring arises from the inclination of the ring, which produces the observed velocity difference between the northwestern and southeastern regions. A sheet-like assembly of the clouds in Cygnus X was already  proposed in \citet{Schneider2023}, based on the overall emission distribution of CO-dark gas only seen in \CII. In particular the DR21 ridge was found to be formed in such a way, demonstrated by the typical observed V shape in \CII-PV cuts \citep{Bonne2023}. This scenario is further supported by dedicated simulations (Sect.~\ref{sec:sim_setup}) and discussed in Sect.~\ref{sec:discuss}. A tilt angle of $\sim$34$^{\rm o}$ and a radial expansion velocity of $\sim$1.3 km s$^{-1}$ was derived with the methodology and details provided in Appendix~\ref{appendix-ellipse}. A similar expansion velocity was obtained from spectra and position-velocity cuts for the \CII\ bubble of NGC 1977 in Orion, driven by a B1V star \citep{Pabst2020}.

\begin{figure*}[htpb]
    \centering
    \includegraphics[width=16cm]{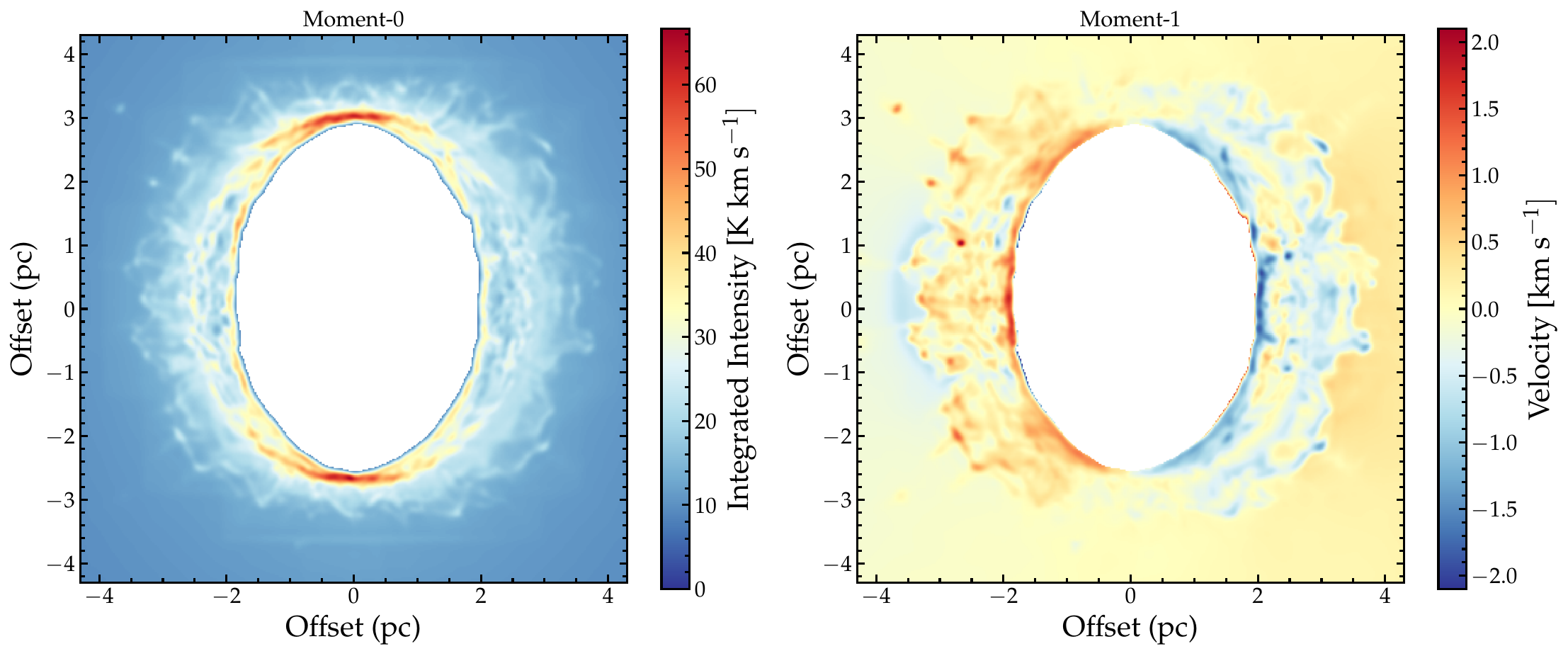}
    \caption{Moment maps of the simulated \CII\ emission at a spatial scale similar to that observed for the Diamond Ring. From the hydrodynamic simulations including stellar wind and with an initial slab density of 350 cm$^{-3}$ and thickness of 2.0 pc, shown in Fig.~\ref{fig:D2_overview}, we applied a 2$\sigma$ intensity threshold based on our observed \CII\ noise level and convolved the maps with a Gaussian kernel of size of 15$''$. The slab in this synthetic observation is rotated by 34$^\circ$ against the plane of the sky to match the observed tilt. The Diamond Ring is additionally rotated by 29.6$^\circ$ counter-clockwise (Fig.\ref{fig:ellipse_moment1}), which is not accounted for here as it does not change the observed LOS velocities. 
    }
    \label{fig:simualted-moment-maps}
\end{figure*}

\subsection{The \HII\ region and the exciting star of the Diamond Ring} \label{subsec:stars}

\citet{Comeron2008} identified 96 candidate massive stars in the outskirts of Cyg~OB2 of which star \#227 is the most likely exciting source for the \HII\ region associated with the Diamond Ring, due to its spatial location. Figure \ref{fig:hii-region} shows the extent of the \HII\ region within the cavity of the Diamond Ring in 24 $\mu$m and 6~cm free-free emission. 
From the spectroscopy of the star (details in Appendix \ref{appendix-spectra}), we conclude that star \#227 is an early-type Be star, most likely a Herbig B0.5e with an uncertainty of about one spectral subclass. These relatively young stars have masses between 10 to 16 M$_\sun$.
\subsection{Physical properties of the Diamond Ring}
\label{sec:physical_properties}
In the following, we only give a summary of the calculated values, with further details given in  Appendix~\ref{appendix-physicalproperties}. 
From the observed line-integrated ($-$8 to 3 km s$^{-1}$) \CII\ intensities and assuming optically thin emission \footnote{We found no indications for a high optical depth or self-absorbed \CII\ line in the DR as it was observed in other PDRs \citep{Kabanovic2022}.}, 
we derived an average column \CII\ density of $N_{\rm CII}=9.7\cdot10^{17}$ cm$^{-2}$, which translates to a total mass of the atomic PDR gas of 1083~M$_\odot$ for the whole ring. From the GLOSTAR cm-data, we obtained the ionising flux and determined an electron density of 86~cm$^{-3}$ for a temperature of 8000~K. \\
By estimating the pressure of the \HII\ region and comparing it to the simulations of \HII\ regions by \citet{Tremblin2014b}, which account for the ram pressure of an external medium, we obtained the picture of an evolved \HII\ region with an age of approximately 2–2.5 \unit{Myr}. The upper limit for the age of the associated \CII\ ring is given by the dynamical timescale inferred from the expansion velocity, assumed to be constant over time,  yielding $t_{\text{dyn}}\sim2.2$ \unit{Myr}. In Sect.~\ref{sec:timescales}, we discuss how these commonly employed age estimates overestimate the age of the observed structure by not accounting for the effect of the environment on the evolution. 

\begin{figure}[htbp]
\begin{center}
\includegraphics[width=8.7cm, trim={0cm 0 0cm 0}, clip, angle={0}]{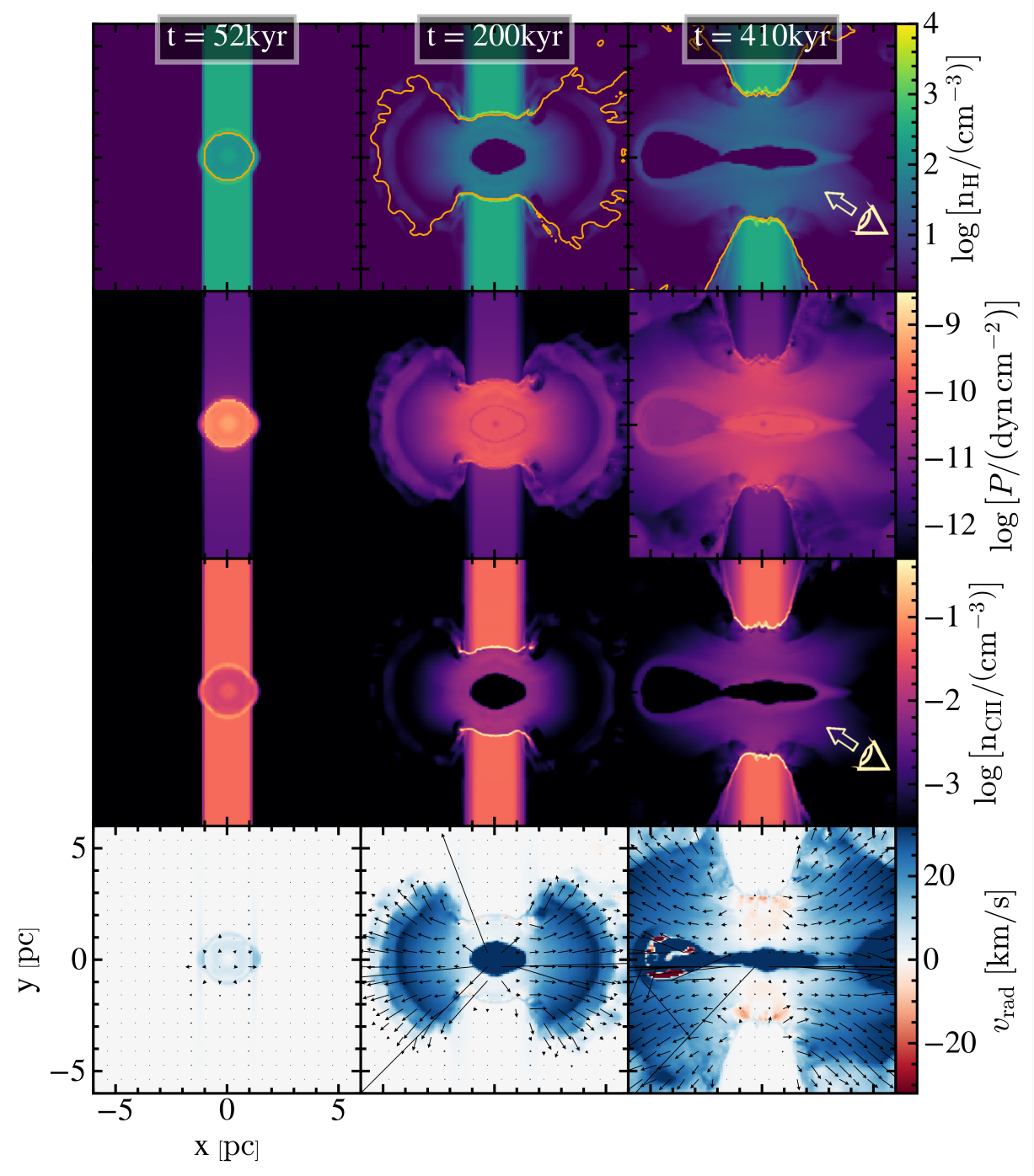}
\vspace{0.0cm}
\caption{
Slices of the central $12\,\mathrm{pc} \times 12\,\mathrm{pc}$ of a simulation with stellar wind of the total hydrogen number density, pressure, \CII\ number density, and radial velocity for three time steps. The full simulation box is further elongated along the x-direction to capture the propagation of the fast bipolar outflow. The time evolution is shown from left to right. The orange contour marks the position of the ionization front and the black arrows show the direction of the gas velocity. The direction with an inclination of 34$^\circ$ from which the DR is observed is indicated in two panels. An animated version of the first and third panel can be found online.
}
\label{fig:D2_overview}
\end{center}
\end{figure}

\subsection{Simulating stellar feedback in a slab of gas}
\label{sec:sim_setup}
The observations discussed in Sect.~\ref{sec:results} reveal an expanding ring of gas traced by the \CII\ emission line, without evidence of a 3D bubble structure typically observed and reported in the literature. This configuration may arise when a star evolves within a `flat' molecular cloud, as proposed by \citet{Beaumont2010}.
To show that such a geometry, combined with stellar feedback effects of a B0.5 star, can explain the observed structure, we performed dedicated simulations. For that purpose, we placed such a star in a slab of gas with varying slab thicknesses and initial densities and conducted simulations with ionising radiation and with and without stellar wind. Trying to use the simplest model that reproduces key observables, we chose to neglect magnetic fields and turbulence.
In Fig. \ref{fig:simualted-moment-maps}, we present synthetic \CII\ observations of the simulation that offer the best match to the observations, but first we begin with a description of the simulation set-up.

We performed 3D hydrodynamics simulations with the Eulerian adaptive mesh refinement (AMR) code \textsc{Flash 4} \citep{Fryxell2000}. The standard piecewise parabolic method (PPM) was used to solve the Euler equations. Self-gravity was calculated using the Octtree-based solver described in \citet{Wunsch2018}. The same solver was used to calculate the local attenuation of the FUV radiation for every cell with the \textsc{TreeRay/OpticalDepth} module. The radiative transport equation for extreme ultraviolet radiation (EUV) was solved with the \textsc{TreeRay/OnTheSpot} module \citep{Wunsch2021}, which uses the on-the-spot approximation in combination with a backwards ray-tracing method. It also utilises the underlying tree structure to make the computation more efficient. As multipole acceptance criterion (MAC) for all \textsc{TreeRay} modules, we used a simple geometric MAC \citep{Barnes1986}, which accepts a node if its angular size is smaller than a chosen limit angle, and set $\theta_\mathrm{lim} = 0.25\,\mathrm{rad}$ (see \citet{Wunsch2018,Wunsch2021} for more details).
We used the chemical network originally implemented in \citet{Glover2007a,Glover2007b} and extended it in \citet{Glover2010} to the network of \citet{Nelson1999}. The implementation in \textsc{Flash 4} was presented in \citet{Walch2015b} and it has since been updated by \citet{Mackey2019} and \citet{Gaches2023}. The network follows the abundances of 32 species. The full non-equilibrium evolution was
followed for 19 species: e$^-$, H$^+$, H, H$_2$, He, He$^+$, C, C$^+$, O, O$^+$, OH, H$_2$O, CO, C$_2$, O$_2$, HCO$^+$, CH, CH$_2$, and CH$_3^+$. The other 13 species were always assumed
to be in chemical equilibrium:\ H$^-$, H$^+_2$, H$^+_3$, CH$^+$, CH$^+_2$, OH$^+$, H$_2$O$^+$, H$_3$O$^+$, CO$^+$, HOC$^+$, O$^-$, C$^-$ and O$_2^+$. The ionising radiation was coupled to the chemistry network, from which the heating and cooling terms were calculated \citep{Haid2018}. The stellar properties were obtained from the stellar tracks of \citet{Ekstroem2012} (hereafter, `Geneva tracks'). The terminal wind velocities were not given in the tracks; therefore, they were calculated according to the procedure described in \citet{Gatto2017}. FUV radiation is treated `adaptively' as described in \citet{Rathjen2025} with some adjustments for this simulation set-up. Instead of using a constant background value, we calculate the FUV luminosity for each star and map the resulting FUV radiation field onto the grid. This takes into account that the radiation field around each star is geometrically diluted according to the inverse-square law ($G_0 \propto R^{-2}$). The FUV luminosity of a star is calculated by integrating its spectrum from $5.6\,\mathrm{eV}$ to $13.6\,\mathrm{eV}$. We assumed the spectrum to be a black body with an effective temperature obtained from the Geneva stellar tracks. A background field of $G_\mathrm{bg} = 0.0948$ \citep{Rathjen2025} is added and the dust attenuation calculated by \textsc{TreeRay/OpticalDepth} is applied.
\textsc{TreeRay/OpticalDepth} was originally designed to simulate a molecular cloud that gets irradiated from the outside by an isotropic background radiation field. It computes the attenuation of FUV radiation as it travels to a point, which we  refer to as the target cell, inside of the MC. The module calculates column densities along multiple sight lines from the perspective of the target cell and averages them to determine the overall attenuation. In our set-up, however, the FUV radiation originates from a single stellar source and it is particularly important to accurately model the FUV field at the PDR, where most of the \CII\ emission is produced, as this determines how much carbon is present in the form of CO, C$^{0}$, or C$^+$. The PDR is located at the edge of the \HII\ region, where the expanding shock continuously sweeps up gas into a swept-up ring (SUR). We assumed that the FUV radiation from the central star is primarily attenuated by the SUR itself. Therefore, we limited the path length over which \textsc{TreeRay/OpticalDepth} integrated column densities along the sight lines to the thickness of the SUR. Unlike the study done by \citet{Rathjen2025}, we did not limit the range of the FUV radiation.

The set-up consisted of an elongated box with dimensions $36\,\mathrm{pc} \times 12 \,\mathrm{pc}\times12\,\mathrm{pc}$ filled with a uniform diffuse gas with a number density of $\mathrm{n}_\mathrm{H}=1\,\mathrm{cm}^{-3}$ and a slab with enhanced density located in the y-z-plane. All gas was initialised in thermal and chemical equilibrium. We varied the slab thickness using values of $1\,$pc, $2\,$pc, $2.5\,$pc, and $3\,$pc and the slab number density using values of $\mathrm{n}_\mathrm{H} = 167 \,\mathrm{cm}^{-3}$, 250 $\mathrm{cm}^{-3}$, 350 $\mathrm{cm}^{-3}$, and 500 $\mathrm{cm}^{-3}$. 
The density range was motivated by the study by \citet{Schneider2023}, where it was shown that molecular clouds built up by the interaction of atomic (number density of at least 100 cm$^{-3}$) and molecular flows at a temperature of $\sim$100 K.

The carbon abundance is set to C/H=1.6$\cdot$10$^{-4}$ \citep{Sofia2004}. All carbon is initialised in the form of C$^+$, which is close to the chemical equilibrium state for all initial densities. The slab consists of mostly H$_2$ with number fractions of above H$_2/\mathrm{H}=0.495$ and temperatures of $T \approx 100\,\mathrm{K}$, where the exact equilibrium abundances and temperatures depend on the slab density. The diffuse surrounding gas has a temperature of $T=814.5\,\mathrm{K}$ and consists of mostly atomic hydrogen with small H$_2/\mathrm{H}=7 \cdot 10^{-3}$ and \HII$/\mathrm{H}=1\cdot10^{-7}$ number fractions.
We used the AMR technique \citep{Fryxell2000} to refine the grid at the position of cells with more than twice the initial slab density, ensuring that the dense and shocked material at the edge of the \HII\ region in the SUR was optimally resolved. The base grid has a resolution of $dx=12\,\mathrm{pc}/128 \approx0.094\,\mathrm{pc}$ and the highest resolution level corresponds to  $dx=12\,\mathrm{pc}/512 \approx0.023\,\mathrm{pc}$.

At $t=0,$ a single massive star was placed in the centre of the slab, creating an \HII\ region initially confined by the dense gas. As the central star, we chose a $16\,\mathrm{M}_\odot$ B0.5 ZAMS star, in accordance with our observations. We simulated its ionising radiation and stellar wind. We speculate that this star was either formed as a result of the interaction of gas flows \citep{Schneider2023}, which built up a dense clump in which a massive star formed, or it is a runaway star that was formed further away and now crosses the slab of gas. 
The mass loss rate of the star is $\mathrm{\dot{M}}_\star = 1.2\cdot 10^{-8}\,\mathrm{M_{\odot}}\mathrm{yr}^{-1}$, the Lyman continuum flux 
\footnote{We derived a slightly higher Lyman flux ($N_{Ly}=1.50\cdot10^{48}$~s$^{-1}$) from GLOSTAR cm data, likely due to EUV photon leakage from nearby ionising stars.} is $N_{Ly}~=~9.8\cdot 10^{47}\,\mathrm{s}^{-1}$, the terminal wind velocity is $\mathrm{v}_w = 2.7\cdot10^{3}\,\mathrm{km} \,\mathrm{s}^{-1}$ and the FUV luminosity is $L_\mathrm{FUV}= 4.6 \cdot 10^{37}\,\mathrm{erg}\,\mathrm{s}^{-1}$.
We ran a set of simulations but show here only the set-up that best matches the observations. We determined the best-matching simulation by comparing integrated intensities, morphology, dynamics, and \HII\ region pressure at a time step when the observed and simulated \CII\ rings matched in size (with a diameter of approximately $6\,$pc). \\
The different evolutionary stages of the expansion of the \HII\ region can be explored in the overview plot (Fig.~\ref{fig:D2_overview}). We discuss the detailed evolution in Sect. \ref{sec:sim_scenario}.
For each simulation, we picked a time step for which the diameter of the slab cavity is around $6\,$pc (the observed size of the Diamond Ring) and perform synthetic observations with RADMC-3D \citep{Dullemond2012} of the $158\,\mu\mathrm{m}$ \CII\ line. The RADMC-3D routine creates a PPV-cube with a velocity range from $-$30 km s$^{-1}$ to 30 km s$^{-1}$ and a resolution of 0.1 km s$^{-1}$. \\
To make a direct comparison with the observations of the Diamond Ring, we resampled the PPV-cubes to 0.3 km s$^{-1}$ and spatially smooth with a Gaussian beam of 15$''$. For the moment-0 and moment-1 maps, we applied a noise mask based on the integrated intensity noise level of the full cube, $I_{\mathrm{rms}}~=\mathrm{rms}_{\mathrm{ch}} \Delta v \sqrt{N_{\mathrm{ch}}}$=4.45 K km s$^{-1}$, where $\mathrm{rms}_{\mathrm{ch}}=1.05$ K is the channel-wise rms noise present in our SOFIA observations at 0.3 km s$^{-1}$ resolution. Based on this noise level, we applied a 2$\sigma$ cut; however, note here that no low-intensity central emission is visible, even up to 1.6$\sigma$.
The structure is viewed with a tilt of 34$^\circ$, with the viewing angle determined by fitting an ellipse to the Diamond Ring as shown in Appendix \ref{appendix-ellipse}.
The resulting moment-0 and moment-1 maps are shown in Fig.~\ref{fig:simualted-moment-maps} and their comparison to the observations will be discussed in the next section.
\section{The evolution of the Diamond Ring} \label{sec:discuss}
\subsection{Observational scenario}
\label{sec:evolution_dr}
For the B0.5e star close to the centre of the Diamond Ring, we would expect moderate mass-loss rates and a lower terminal wind velocity compared to more massive O-type stars. Nevertheless, the thermal pressure of the ionising radiation and thermalisation of the wind's kinetic energy is expected to create a warm ($\sim$8000~K) bubble of ionised gas that drives expansion velocities of 1-10 km s$^{-1}$, 
consistent with previously observed \CII\ shell velocities \citep{Pabst2019,Luisi2021}. 
 Here, we argue that the Diamond Ring initially formed in a slab of gas. After an initial expansion driven by the pressure of the \HII\ region, the shell will break out at the bottom and top of the slab and ionised gas will escape into the lower density environment. The radial expansion in the plane of the slab will continue to be driven by the momentum impacted by the ionised gas. As it sweeps up more and more mass, the velocity of this radial expansion will decrease. 
The ionising feedback of the star creates a bipolar outflow perpendicular to the layer and a shock front sweeps up molecular gas within the layer, forming a dense ring confined to the plane of the molecular slab.
The 24 $\mu$m emission  (Fig.~\ref{fig:hii-region} a) centrally peaks at the position of star \#227, while the cm emission (Fig.~\ref{fig:hii-region} c) is brighter to the east of the star. This may reflect a dust temperature gradient driven by stellar heating, with warmer dust near the central star naturally producing enhanced 24 $\mu$m emission \citep{Tielens2005}. The idea that an \HII\ region (with its associated \CII\ bubble) bursts out of a `flat' molecular cloud or a slab is supported by several observational studies. For instance, bipolar \HII\ regions have been observed in RCW36 \citep{Bonne2022} and the Rosette nebula \citep{Wareing2018}, potentially driven by the feedback of one ionizing O star (five O-type stars in the latter case) in a molecular slab.
Flattened molecular cloud structures around \HII\ regions were also  observed in the J~=~3$\to$2 line transition of CO in the work of \cite{Beaumont2010}. The molecular gas distribution appears ring-like (not shell-like) in all studied objects (i.e. 43 {\sl Spitzer}-selected bubbles). A LOS CO emission deficit was also observed in RCW120 by \cite{Kabanovic2022}, where the authors argue that the observed limb-brightening effect and molecular line widths cannot be reproduced by a spherically symmetric molecular cloud. Moreover, the initial structure of the associated molecular cloud must have been flat. \citet{Schneider2023} showed that low surface brightness \CII\ intensity in the Cygnus X region (including the DR21 ridge, the W75N star-forming regions,  and the Diamond Ring area) arises from CO-dark molecular and atomic gas. The observed intensities can be explained by interacting flows of gas that have a filamentary or slab-like geometry. When combined with slightly higher density pre-existing molecular clumps and the filamentary connection to the DR21 ridge in the north-west, this scenario provides a plausible explanation for the observed patchy CO emission.
In addition, in the SILCC-Zoom simulations presented in  \citet{Ganguly2023}, molecular clouds exhibit predominantly sheet-like morphologies, with spheroidal structures being rare. 
In our simplified simulation set-up, we did not try to reproduce the global dynamics of the gas; therefore, we could not accurately predict the physical properties of  molecular gas outside the reach of the ring.

\subsection{Simulation scenario}
\label{sec:sim_scenario}

Comparing the observed intensities and dynamics with the simulations enabled us to better constrain the parameter space in our scenario. 
We find that a slab with 2.0~pc thickness and a hydrogen density of $n_{\rm H}$=350 cm$^{-3}$ gives the best match, as illustrated in the moment-maps of Fig.~\ref{fig:simualted-moment-maps}, obtained from the original simulations displayed in Fig.~\ref{fig:D2_overview}. Simulations with thinner slabs (e.g. $1\,$pc) compress the SUR gas into a small region. The high-density gas then forms CO and has a lower C$^+$ fraction. This is incompatible with the DR, because the \CII\ intensities are too low and CO would be detectable in the SUR. Simulations with thicker slabs (e.g. $2.5\,$pc and $3\,$pc) show a different morphology in the moment maps. Multiple overlaying rings are visible and the central void tapers to a point at the ends of the long axis. The intensity at these two points is also strongly increased. 

\begin{figure}[htbp]
\begin{center}
\includegraphics[width=8.7cm, trim={0cm 0 0cm 0}, clip, angle={0}]{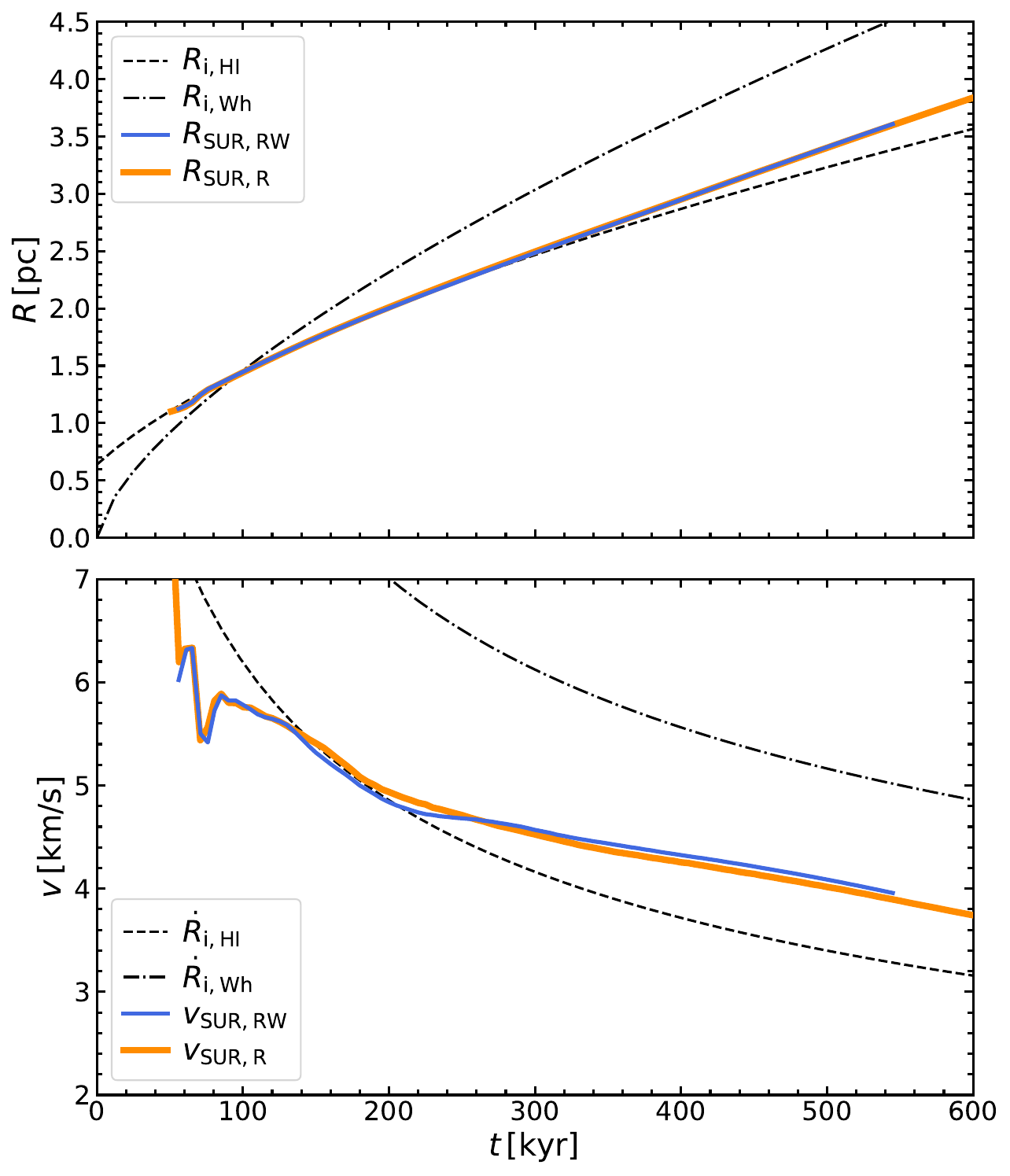}
\vspace{0.0cm}
\caption{
Evolution of radius and radial velocity of the SUR for simulations with ionising radiation and wind, \texttt{RW}, and with only ionising radiation, \texttt{R}. The analytical solutions for the Hosokawa-Inutsuka (HI) and Whitworth (Wh) models are plotted for comparison.}
\label{fig:radius_and_velocity}
\end{center}
\end{figure}

Based on Fig. \ref{fig:D2_overview}, we can explain the evolution of the best-fit simulation. The \HII\ region is initially confined to the slab and evolves according to the analytical solution by \citet{Hosokawa2006} (see Eq. \ref{eq:HoIn}). Then, at around $100\,\mathrm{kyr}$, it breaks out of the slab and produces an outflow led by two half shells. In the second snapshot, the outflowing gas is more diffuse and fully photo-ionised and the ionising radiation escapes in a bipolar `champagne flow' \citep{Tenorio-Tagle1979}. At this stage, the expansion of the \HII\ region into the slab transitions to the scenario analysed by \citet{Whitworth2022} of a star eroding a molecular slab. An inwardly directed photo-evaporation flow develops, which (due to the conservation of momentum) must accelerate the SUR outwards. This effect is known as the `rocket effect', first proposed by \citet{Oort1954}. Thus, even though the pressure in the bubble decreases and, along with it, its acceleration on the SUR, the `rocket effect' leads to a continued expansion. The last snapshot shows the simulation at the time the SUR reaches the size of the DR with a diameter of $6\,\mathrm{pc}$. The two half shells have left thedisplayed area, but are still inside of the elongated simulation box. The direction from which the DR is observed from is indicated in the first and the third panel. The stellar wind creates an elongated bubble that expands primarily along the direction of the outflow. The bubble is filled with very hot ($T>10^{7}\,$K), low-density ($n < 10^{-2}\,\mathrm{cm}^{-3}$) gas. At this stage, the SUR contains about $0.5\,\mathrm{M}_\odot$ of C$^+$ and has a mean number density of $n_\mathrm{SUR}\sim3000\,\mathrm{cm^{-3}}$.
Figure \ref{fig:radius_and_velocity} shows the evolution of the SUR radius and its radial velocity. We selected the gas within the SUR by applying a density threshold of twice the initial slab density and filtering for a positive radial gas velocity. We put the simulation results into perspective by comparing the size and velocity evolution with two analytical models, which describe the expansion of an \HII\ region for two different set-ups, but which do not include the effects of stellar winds. In the first phase, the expansion of the ionisation front (IF) follows the solution of an \HII\ region expanding in a uniform medium,
\begin{equation} \label{eq:HoIn}
    R_\mathrm{i,HI}(t) = R_\mathrm{St} \left(1 + \frac{7}{4}\sqrt{\frac{4}{3}}\frac{C_{\mathrm{H_{II}}} t}{R_\mathrm{St}} \right)^{4/7},
\end{equation}
which was derived by \citet{Hosokawa2006}  from the equation of motion of the shell. Here, $R_\mathrm{St}$ is the Strömgren radius  and $C_\mathrm{H_{II}}$ the isothermal sound speed of the ionised gas. At around $150\,$kyr, the region transforms into a bipolar \HII\ region, for which \citet{Whitworth2022} derived the expansion rate of the IF, expressed as
\begin{multline}\label{eq:Whitworth}
    R_\mathrm{i,Wh}(t) \sim3.8\,\mathrm{pc}\left[\frac{\mathrm{Z}}{0.2\,\mathrm{pc}}\right]^{-1/6}\left[\frac{n_\mathrm{H_2}}{10^4\,\mathrm{cm^{-3}}}\right]^{-1/3} \\
    \left[ \frac{N_\mathrm{Ly}}{10^{49}\,\mathrm{s}^{-1}}\right]^{1/6} \left[ \frac{t}{\mathrm{1\,Myr}} \right]^{2/3},
\end{multline}
with Z being the slab thickness and $n_\mathrm{H_2}$ the H$_2$ number density of the slab, which is assumed to be fully molecular. Therefore, the H$_2$ number density in Equation \ref{eq:Whitworth} is related to the total hydrogen number densities used in our simulations simply by $n_\mathrm{H_2} = 0.5\,n_\mathrm{H}$. By taking the time derivative of both equations, we find that the velocities with which the IF expands scale as $\dot{R}_\mathrm{i,HI}(t)\propto t^{-3/7}$, for $t \gg t_\mathrm{dyn} = R_\mathrm{St}/C_\mathrm{H_{II}}$, and $\dot{R}_\mathrm{i,Wh}(t) \propto t^{-1/3}$. The expansion velocity in the \citet{Hosokawa2006} solution decreases more slowly so that a larger expansion velocity can be sustained for longer. We note that while $\dot{R}_\mathrm{i}$ is the velocity of the IF and not the gas velocity of the gas in the SUR, they are approximately equal. This is because, compared to the mass in the SUR, only a small amount of gas gets ionised and flows into the \HII\ region. We find that the stellar wind does not significantly affect the size and velocity of the SUR and only increases its radial momentum by $1.6\,\%$.

In our simulation set-up, the expansion is slowly decreasing with diameter after the initial breakout. The expanding ring reaches a diameter of size $\sim$6 pc after $400-500\,\mathrm{kyr}$ (Fig.~\ref{fig:radius_and_velocity}), earlier than the classical age estimates that assume spherical symmetry. 
Furthermore, the velocity dispersion and the order of magnitude of the pressure of the \HII\ region are reproduced by our simple simulation picture (Fig.~\ref{fig:D2_overview} and Fig.~\ref{fig:ellipse_moment2}). 

The most notable distinction between observations and simulations is the difference in the \CII\ integrated intensity. We assume that the higher \CII\ intensities in the northeastern part of the ring are driven by mass infall via filaments on the clump containing  protostars, creating a more massive PDR compared to the rest of the ring.
Excluding this region and considering the observational noise for the synthetic observations, the integrated intensities are a good match. Small differences of lower intensities ($\sim$30\% decrease) in our simulations can arise due to clumpiness and also from the limited numerical resolution at the position of the PDR \citep{Franeck2018}. In general, introducing clumpiness can increase the \CII\ emission because FUV photons would penetrate further into the ISM through low-density channels. A clumpy PDR would give a larger surface area, resulting in broadened layering within the PDR (see e.g. \citep{Roellig2022}). For the simulations, we expect our FUV approximation at the PDR to be quite precise. For our set-up a substantial boost of the \CII\ line intensity from a clumpy medium is unlikely because with a low average $A_{V, 3D}\sim0.44$ in the PDR and a mean density of $n_\mathrm{SUR}\sim3000\,\mathrm{cm^{-3}}$, we are already above the critical density and the upper level population is saturated (see e.g., appendix in \citet{Bisbas_2022}). Both analytical models, namely, those of \citet{Hosokawa2006} and \citet{Whitworth2022} and ours, predict gas expansion velocities of $\sim$4 km~s$^{-1}$, which is a factor of 2 higher than determined from the observed LOS velocity difference and the ellipse fit. Such discrepancies are not unexpected in an idealised set-up and we suggest two possible explanations that may help reconcile the tension. One possibility is the influence of a magnetic field, which was not included in our simulations. A field oriented perpendicular to the slab would slow down the radial expansion \citep{Arthur_2011}. For our configuration, a plausible magnetic field strength of $\sim 40\,\mu$G \citep{Crutcher2010} would yield a magnetic pressure comparable to the thermal pressure of the \HII\ region at $t = 400\,$kyr, thereby substantially reducing the expansion rate. The second possible explanation is that in our uniform model, we do not model the dense core in which the star would have originally formed. The core would provide additional gas mass which the feedback needs to accelerate, thereby also slowing down the expansion.

The slab thickness is probably the most important parameter that influences the simulations:  
A larger slab thickness increases the \CII\ intensity and the width of the ring along the minor axis (because there is a larger projected ionising front), while keeping the density constant. 
Furthermore, increasing the slab thickness alters the geometry, making it more bubble-like rather than purely ring-like. As a result, material located closer to the central star in projection can contribute to the LOS velocity, leading to the observation of higher velocities.
We find that slab thicknesses up to 2.5 \unit{pc} reproduce well the observed velocity structure 
(Fig.~\ref{fig:simualted-moment-maps}). 
Based on the noise-level in our observations, we know that we are sensitive from $N_{\rm CII}=3.5\cdot10^{16}$ cm$^{-2}$ upward in our moment-0 map (Fig.\ref{fig:ellipse_moment1}). Applying such a threshold to our observations, it becomes clear that the free-streaming high-velocity+low-density gas component along the LOS in the simulations is below our detection limit, giving rise to the starkly ring-like appearance. 
\subsection{Timescales}
\label{sec:timescales}
We observed a mismatch between the age estimates of around 2~Myr of the \HII\ region derived from the simulations of \citet{Tremblin2014b}, described in Appendix \ref{appendix-physicalproperties}, and the  dynamical age estimate (upper limit), respectively, along with the first appearance of a ring of \CII\ emission in our simulations (typically around 100 kyr, depending on model set-up).
In \cite{Tremblin2014b}, the pressure in the bubble would constantly decrease with $r^{-3/2}$ for most of its evolution as long as the pressure of the ionised gas is higher than the pressure of its surrounding medium, in which it is confined. This leads to an over-estimation of the \HII\ region age by a factor of $\sim5-6$ based on gas pressure for evolved regions in flat molecular clouds. Our findings of bubble destruction starting in the first few 100~kyr are in line with timescales on which cloud dispersion starts to destroy the bubble geometry \citep{Bonne2023b} and well below the expected life time of a B0-type star of $\sim$10~Myr. Introducing clumpiness into the molecular slab could further tighten or relax the tension \citep{Walch2012}. For example, a high fractal dimension would lead to photons leak into the thinner parts and a faster advance of the \HII\ region in certain regions, with pillars protruding into the \HII\ region. This would further decrease the age at which the ring reaches the observed physical size. The fact that we observe a slow radial expansions and no pillars, however, hints at a low initial clumpiness in the molecular slab around the ring.

\section{Summary \& conclusions}   \label{sec:summary}
In this study, we present the first detection of an apparent ring of \CII\ emission, called the Diamond Ring, in the Cygnus~X region that shows only a slow radial expansion of $\sim$1.3 km s$^{-1}$ in the plane of the sky. The Diamond Ring does not display the 3D shell structure typically observed in other \CII\ bubbles.
\\ \\
\noindent $\bullet$ The ring has an atomic mass of $\sim$10$^3$ M$_{\odot}$ and bulk emission at a LOS velocity around $-$2 km s$^{-1}$. \\ \\
\noindent $\bullet$ The associated \HII\ region is ionised by a B0.5e star, as determined by optical spectroscopy. \\ \\
\noindent $\bullet$ The `Diamond' in the Diamond Ring is an unrelated star forming gas clump at a different velocity ($\sim7$ km s$^{-1}$) and contains a small cluster of stars (Cl13 from \citealt{LeDuigou2002}). \\ \\
\noindent $\bullet$ Dedicated 3D simulations confirm that the observed properties can be best explained by a scenario in which the \HII\ region initially formed within a $\sim$2~pc thick slab with a density of $\sim$350~cm$^{-3}$. 
The parts of the shell moving perpendicular to the slab get rapidly diluted in the lower-density surroundings, so that only the slowly expanding ring remains detectable, which is confined by the swept-up gas in the slab plane. \\ \\
\noindent $\bullet$ From the simulations, the age of the Diamond Ring is determined to be around 425~kyr, with a bubble-like expansion only in the first 100~kyr. This is in tension with the commonly employed dynamical age estimate assuming a constant expansion velocity and with \HII\ region age estimates based on turbulent pressure simulations (both methods arrive at $\sim$2 Myrs).
\\ \\
In summary, we propose  we have detected the terminal phase of evolution for expanding \CII\ bubbles within flat molecular clouds or slabs driven by thermal pressure and stellar winds. This study offers insights into how stellar feedback of a massive star impacts the  dynamics of the gas in slab-like molecular environments, which are likely to be more common in star-forming regions than the idealised spherical expansion scenarios that  are typically modelled.

\begin{acknowledgements}
This study was based on observations made with the NASA/DLR
Stratospheric Observatory for Infrared Astronomy (SOFIA). SOFIA is jointly operated by the Universities Space Research Association Inc. (USRA), under NASA contract NNA17BF53C, and the Deutsches SOFIA Institut (DSI), under DLR contract 50OK0901 to the University of Stuttgart. upGREAT is a development by the MPIfR 
and the KOSMA/Universit\"at zu K\"oln, in cooperation with the DLR Institut f\"ur Optische Sensorsysteme.
This work makes use of observations made with the Nordic Optical Telescope, owned in collaboration by the University of Turku and Aarhus University, and operated jointly by Aarhus University, the University of Turku and the University of Oslo, representing Denmark, Finland and Norway, the University of Iceland and Stockholm University at the Observatorio del Roque de los Muchachos, La Palma, Spain, of the Instituto de Astrofisica de Canarias. The software used in this work was developed in part by the DOE NNSA- and DOE Office of Science-supported Flash Center for Computational Science at the University of Chicago and the University of Rochester. The GLOSTAR data product shown here are the results of observations with the Effelsberg 100 meter radio telescope operated by the Max Planck Institute for Radio Astronomy (MPIfR).
Financial support for the SOFIA Legacy Program FEEDBACK at the University of Maryland was provided by NASA through award SOF070077 issued by USRA. 
The FEEDBACK-plus project that is supported by the BMWI via DLR, Projekt Number 50OR2217.  
S.K. acknowledges support by the BMWI via DLR, project number 50OR2311. 
This work was supported by the CRC1601 (SFB 1601 sub-project A6, B1, B2, B4, B6, B7, C1, C2, C3, C6, and C7) funded by the DFG (German Research Foundation) – 500700252.
S.V., S.W., and D.S. acknowledge support by the Ministry of Culture and Science of the State of North Rhine-Westphalia by the Profilbildung project "Big Bang to Big Data" (B3D).
YG was supported by the Ministry of Science and Technology of China under the National Key R\&D Program (Grant No. 2023YFA1608200) and the National Natural Science Foundation of China (Grant No. 12427901).
S.D. acknowledges support by the International Max Planck Research School (IMPRS) for Astronomy and Astrophysics at the Universities of Bonn and Cologne. \\
\end{acknowledgements}
\bibliographystyle{aa} 
\bibliography{bibliography} 

\begin{thebibliography}{74}
\expandafter\ifx\csname natexlab\endcsname\relax\def\natexlab#1{#1}\fi

\bibitem[{Arthur {et~al.}(2011)Arthur, Henney, Mellema, De~Colle, \& Vázquez-Semadeni}]{Arthur_2011}
Arthur, S.~J., Henney, W.~J., Mellema, G., De~Colle, F., \& Vázquez-Semadeni, E. 2011, Monthly Notices of the Royal Astronomical Society, 414, 1747

\bibitem[{{Barnes} \& {Hut}(1986)}]{Barnes1986}
{Barnes}, J. \& {Hut}, P. 1986, \nat, 324, 446

\bibitem[{{Beaumont} \& {Williams}(2010)}]{Beaumont2010}
{Beaumont}, C.~N. \& {Williams}, J.~P. 2010, \apj, 709, 791

\bibitem[{{Beerer} {et~al.}(2010){Beerer}, {Koenig}, {Hora}, {Gutermuth}, {Bontemps}, {Megeath}, {Schneider}, {Motte}, {Carey}, {Simon}, {Keto}, {Smith}, {Allen}, {Fazio}, {Kraemer}, {Price}, {Mizuno}, {Adams}, {Hern{\'a}ndez}, \& {Lucas}}]{Beerer2010}
{Beerer}, I.~M., {Koenig}, X.~P., {Hora}, J.~L., {et~al.} 2010, \apj, 720, 679

\bibitem[{{Beuther} {et~al.}(2022){Beuther}, {Schneider}, {Simon}, {Suri}, {Ossenkopf-Okada}, {Kabanovic}, {R{\"o}llig}, {Guevara}, {Tielens}, {Sandell}, {Buchbender}, {Ricken}, \& {G{\"u}sten}}]{Beuther2022}
{Beuther}, H., {Schneider}, N., {Simon}, R., {et~al.} 2022, \aap, 659, A77

\bibitem[{Bisbas {et~al.}(2022)Bisbas, Walch, Naab, Lahén, Herrera-Camus, Steinwandel, Fotopoulou, Hu, \& Johansson}]{Bisbas_2022}
Bisbas, T.~G., Walch, S., Naab, T., {et~al.} 2022, The Astrophysical Journal, 934, 115

\bibitem[{{Bonne} {et~al.}(2023){Bonne}, {Bontemps}, {Schneider}, {Simon}, {Clarke}, {Csengeri}, {Chambers}, {Graf}, {Jackson}, {Klein}, {Okada}, {Tielens}, \& {Tiwari}}]{Bonne2023}
{Bonne}, L., {Bontemps}, S., {Schneider}, N., {et~al.} 2023, \apj, 951, 39

\bibitem[{{Bonne} {et~al.}(2022){Bonne}, {Schneider}, {Garc{\'\i}a}, {Bij}, {Broos}, {Fissel}, {Guesten}, {Jackson}, {Simon}, {Townsley}, {Zavagno}, {Aladro}, {Buchbender}, {Guevara}, {Higgins}, {Jacob}, {Kabanovic}, {Karim}, {Soam}, {Stutzki}, {Tiwari}, {Wyrowski}, \& {Tielens}}]{Bonne2022}
{Bonne}, L., {Schneider}, N., {Garc{\'\i}a}, P., {et~al.} 2022, \apj, 935, 171

\bibitem[{{Bonne, L.} {et~al.}(2023){Bonne, L.}, {Kabanovic, S.}, {Schneider, N.}, {Zavagno, A.}, {Keilmann, E.}, {Simon, R.}, {Buchbender, C.}, {Güsten, R.}, {Jacob, A. M.}, {Jacobs, K.}, {Kavak, U.}, {Polles, F. L.}, {Tiwari, M.}, {Wyrowski, F.}, \& {Tielens, A. G. G. M.}}]{Bonne2023b}
{Bonne, L.}, {Kabanovic, S.}, {Schneider, N.}, {et~al.} 2023, \aap, 679, L5

\bibitem[{{Brunthaler} {et~al.}(2021){Brunthaler}, {Menten}, {Dzib}, {Cotton}, {Wyrowski}, {Dokara}, {Gong}, {Medina}, {M{\"u}ller}, {Nguyen}, {Ortiz-Le{\'o}n}, {Reich}, {Rugel}, {Urquhart}, {Winkel}, {Yang}, {Beuther}, {Billington}, {Carrasco-Gonzalez}, {Csengeri}, {Murugeshan}, {Pandian}, \& {Roy}}]{Brunthaler2021}
{Brunthaler}, A., {Menten}, K.~M., {Dzib}, S.~A., {et~al.} 2021, \aap, 651, A85

\bibitem[{{Cao} {et~al.}(2019){Cao}, {Qiu}, {Zhang}, {Wang}, {Hu}, \& {Liu}}]{Cao2019}
{Cao}, Y., {Qiu}, K., {Zhang}, Q., {et~al.} 2019, \apjs, 241, 1

\bibitem[{{Carrasco} {et~al.}(2016){Carrasco}, {Evans}, {Montegriffo}, {Jordi}, {van Leeuwen}, {Riello}, {Voss}, {De Angeli}, {Busso}, {Fabricius}, {Cacciari}, {Weiler}, {Pancino}, {Brown}, {Holland}, {Burgess}, {Osborne}, {Altavilla}, {Gebran}, {Ragaini}, {Galleti}, {Cocozza}, {Marinoni}, {Bellazzini}, {Bragaglia}, {Federici}, \& {Balaguer-N{\'u}{\~n}ez}}]{Carrasco2016}
{Carrasco}, J.~M., {Evans}, D.~W., {Montegriffo}, P., {et~al.} 2016, \aap, 595, A7

\bibitem[{{Comer{\'o}n} {et~al.}(2008){Comer{\'o}n}, {Pasquali}, {Figueras}, \& {Torra}}]{Comeron2008}
{Comer{\'o}n}, F., {Pasquali}, A., {Figueras}, F., \& {Torra}, J. 2008, \aap, 486, 453

\bibitem[{Crutcher {et~al.}(2010)Crutcher, Wandelt, Heiles, Falgarone, \& Troland}]{Crutcher2010}
Crutcher, R.~M., Wandelt, B., Heiles, C., Falgarone, E., \& Troland, T.~H. 2010, The Astrophysical Journal, 725, 466

\bibitem[{{Dullemond} {et~al.}(2012){Dullemond}, {Juhasz}, {Pohl}, {Sereshti}, {Shetty}, {Peters}, {Commercon}, \& {Flock}}]{Dullemond2012}
{Dullemond}, C.~P., {Juhasz}, A., {Pohl}, A., {et~al.} 2012

\bibitem[{{Dutra} \& {Bica}(2001)}]{Dutra2001}
{Dutra}, C.~M. \& {Bica}, E. 2001, \aap, 376, 434

\bibitem[{{Ekstr{\"o}m} {et~al.}(2012){Ekstr{\"o}m}, {Georgy}, {Eggenberger}, {Meynet}, {Mowlavi}, {Wyttenbach}, {Granada}, {Decressin}, {Hirschi}, {Frischknecht}, {Charbonnel}, \& {Maeder}}]{Ekstroem2012}
{Ekstr{\"o}m}, S., {Georgy}, C., {Eggenberger}, P., {et~al.} 2012, \aap, 537, A146

\bibitem[{{Franeck} {et~al.}(2018){Franeck}, {Walch}, {Seifried}, {Clarke}, {Ossenkopf-Okada}, {Glover}, {Klessen}, {Girichidis}, {Naab}, {W{\"u}nsch}, {Clark}, {Pellegrini}, \& {Peters}}]{Franeck2018}
{Franeck}, A., {Walch}, S., {Seifried}, D., {et~al.} 2018, \mnras, 481, 4277

\bibitem[{{Fryxell} {et~al.}(2000){Fryxell}, {Olson}, {Ricker}, {Timmes}, {Zingale}, {Lamb}, {MacNeice}, {Rosner}, {Truran}, \& {Tufo}}]{Fryxell2000}
{Fryxell}, B., {Olson}, K., {Ricker}, P., {et~al.} 2000, \apjs, 131, 273

\bibitem[{Gaches {et~al.}(2023)Gaches, Walch, Wünsch, \& Mackey}]{Gaches2023}
Gaches, B. A.~L., Walch, S., Wünsch, R., \& Mackey, J. 2023, Monthly Notices of the Royal Astronomical Society, 522, 4674

\bibitem[{{Ganguly} {et~al.}(2023){Ganguly}, {Walch}, {Seifried}, {Clarke}, \& {Weis}}]{Ganguly2023}
{Ganguly}, S., {Walch}, S., {Seifried}, D., {Clarke}, S.~D., \& {Weis}, M. 2023, \mnras, 525, 721

\bibitem[{{Gatto} {et~al.}(2017){Gatto}, {Walch}, {Naab}, {Girichidis}, {W{\"u}nsch}, {Glover}, {Klessen}, {Clark}, {Peters}, {Derigs}, {Baczynski}, \& {Puls}}]{Gatto2017}
{Gatto}, A., {Walch}, S., {Naab}, T., {et~al.} 2017, \mnras, 466, 1903

\bibitem[{{Glover} {et~al.}(2010){Glover}, {Federrath}, {Mac Low}, \& {Klessen}}]{Glover2010}
{Glover}, S.~C.~O., {Federrath}, C., {Mac Low}, M.~M., \& {Klessen}, R.~S. 2010, \mnras, 404, 2

\bibitem[{{Glover} \& {Mac Low}(2007{\natexlab{a}})}]{Glover2007a}
{Glover}, S. C.~O. \& {Mac Low}, M.-M. 2007{\natexlab{a}}, \apjs, 169, 239

\bibitem[{{Glover} \& {Mac Low}(2007{\natexlab{b}})}]{Glover2007b}
{Glover}, S. C.~O. \& {Mac Low}, M.-M. 2007{\natexlab{b}}, \apj, 659, 1317

\bibitem[{{Goldsmith} {et~al.}(2012){Goldsmith}, {Langer}, {Pineda}, \& {Velusamy}}]{Goldsmith2012}
{Goldsmith}, P.~F., {Langer}, W.~D., {Pineda}, J.~L., \& {Velusamy}, T. 2012, \apjs, 203, 13

\bibitem[{{Gong} {et~al.}(2023){Gong}, {Ortiz-Le{\'o}n}, {Rugel}, {Menten}, {Brunthaler}, {Wyrowski}, {Henkel}, {Beuther}, {Dzib}, {Urquhart}, {Yang}, {Pandian}, {Dokara}, {Veena}, {Nguyen}, {Medina}, {Cotton}, {Reich}, {Winkel}, {M{\"u}ller}, {Skretas}, {Csengeri}, {Khan}, \& {Cheema}}]{Gong2023}
{Gong}, Y., {Ortiz-Le{\'o}n}, G.~N., {Rugel}, M.~R., {et~al.} 2023, \aap, 678, A130

\bibitem[{{Haid} {et~al.}(2018){Haid}, {Walch}, {Seifried}, {W{\"u}nsch}, {Dinnbier}, \& {Naab}}]{Haid2018}
{Haid}, S., {Walch}, S., {Seifried}, D., {et~al.} 2018, \mnras, 478, 4799

\bibitem[{{Hennemann} {et~al.}(2012){Hennemann}, {Motte}, {Schneider}, {Didelon}, {Hill}, {Arzoumanian}, {Bontemps}, {Csengeri}, {Andr{\'e}}, {Konyves}, {Louvet}, {Marston}, {Men'shchikov}, {Minier}, {Nguyen Luong}, {Palmeirim}, {Peretto}, {Sauvage}, {Zavagno}, {Anderson}, {Bernard}, {Di Francesco}, {Elia}, {Li}, {Martin}, {Molinari}, {Pezzuto}, {Russeil}, {Rygl}, {Schisano}, {Spinoglio}, {Sousbie}, {Ward-Thompson}, \& {White}}]{Hennemann2012}
{Hennemann}, M., {Motte}, F., {Schneider}, N., {et~al.} 2012, \aap, 543, L3

\bibitem[{{Hollenbach} \& {Tielens}(1999)}]{Hollenbach1999}
{Hollenbach}, D.~J. \& {Tielens}, A.~G.~G.~M. 1999, Reviews of Modern Physics, 71, 173

\bibitem[{{Hosokawa} \& {Inutsuka}(2006)}]{Hosokawa2006}
{Hosokawa}, T. \& {Inutsuka}, S.-i. 2006, \apj, 646, 240

\bibitem[{{Kabanovic} {et~al.}(2022){Kabanovic}, {Schneider}, {Ossenkopf-Okada}, {Falasca}, {G{\"u}sten}, {Stutzki}, {Simon}, {Buchbender}, {Anderson}, {Bonne}, {Guevara}, {Higgins}, {Koribalski}, {Luisi}, {Mertens}, {Okada}, {R{\"o}llig}, {Seifried}, {Tiwari}, {Wyrowski}, {Zavagno}, \& {Tielens}}]{Kabanovic2022}
{Kabanovic}, S., {Schneider}, N., {Ossenkopf-Okada}, V., {et~al.} 2022, \aap, 659, A36

\bibitem[{{Keilmann} {et~al.}(2025){Keilmann}, {Dannhauer}, {Kabanovic}, {Schneider}, {Ossenkopf-Okada}, {Simon}, {Bonne}, {Goldsmith}, {G{\"u}sten}, {Zavagno}, {Stutzki}, {Riechers}, {R{\"o}llig}, {Verbena}, \& {Tielens}}]{Keilmann2025}
{Keilmann}, E., {Dannhauer}, S., {Kabanovic}, S., {et~al.} 2025, \aap, 697, L2

\bibitem[{{Kryukova} {et~al.}(2014){Kryukova}, {Megeath}, {Hora}, {Gutermuth}, {Bontemps}, {Kraemer}, {Hennemann}, {Schneider}, {Smith}, \& {Motte}}]{Kryukova2014}
{Kryukova}, E., {Megeath}, S.~T., {Hora}, J.~L., {et~al.} 2014, \aj, 148, 11

\bibitem[{{Kuchar} \& {Clark}(1997)}]{Kuchar1997}
{Kuchar}, T.~A. \& {Clark}, F.~O. 1997, \apj, 488, 224

\bibitem[{{Larson}(1981)}]{Larson1981}
{Larson}, R.~B. 1981, \mnras, 194, 809

\bibitem[{{Le Duigou} \& {Kn{\"o}dlseder}(2002)}]{LeDuigou2002}
{Le Duigou}, J.~M. \& {Kn{\"o}dlseder}, J. 2002, \aap, 392, 869

\bibitem[{{Lockman}(1989)}]{Lockman1989}
{Lockman}, F.~J. 1989, \apjs, 71, 469

\bibitem[{{Luisi} {et~al.}(2021){Luisi}, {Anderson}, {Schneider}, {Simon}, {Kabanovic}, {G{\"u}sten}, {Zavagno}, {Broos}, {Buchbender}, {Guevara}, {Jacobs}, {Justen}, {Klein}, {Linville}, {R{\"o}llig}, {Russeil}, {Stutzki}, {Tiwari}, {Townsley}, \& {Tielens}}]{Luisi2021}
{Luisi}, M., {Anderson}, L.~D., {Schneider}, N., {et~al.} 2021, Science Advances, 7, eabe9511

\bibitem[{{Mackey} {et~al.}(2019){Mackey}, {Walch}, {Seifried}, {Glover}, {W{\"u}nsch}, \& {Aharonian}}]{Mackey2019}
{Mackey}, J., {Walch}, S., {Seifried}, D., {et~al.} 2019, \mnras, 486, 1094

\bibitem[{{Marston} {et~al.}(2004){Marston}, {Reach}, {Noriega-Crespo}, {Rho}, {Smith}, {Melnick}, {Fazio}, {Rieke}, {Carey}, {Rebull}, {Muzerolle}, {Egami}, {Watson}, {Pipher}, {Latter}, \& {Stapelfeldt}}]{Marston2004}
{Marston}, A.~P., {Reach}, W.~T., {Noriega-Crespo}, A., {et~al.} 2004, \apjs, 154, 333

\bibitem[{{Mart{\'{\i}}n-Hern{\'a}ndez} {et~al.}(2005){Mart{\'{\i}}n-Hern{\'a}ndez}, {Vermeij}, \& {van der Hulst}}]{MartinHernandez2005}
{Mart{\'{\i}}n-Hern{\'a}ndez}, N.~L., {Vermeij}, R., \& {van der Hulst}, J.~M. 2005, A\&A, 433, 205

\bibitem[{{Nelson} \& {Langer}(1999)}]{Nelson1999}
{Nelson}, R.~P. \& {Langer}, W.~D. 1999, \apj, 524, 923

\bibitem[{{Oort}(1954)}]{Oort1954}
{Oort}, J.~H. 1954, \bain, 12, 177

\bibitem[{{Pabst} {et~al.}(2019){Pabst}, {Higgins}, {Goicoechea}, {Teyssier}, {Berne}, {Chambers}, {Wolfire}, {Suri}, {Guesten}, {Stutzki}, {Graf}, {Risacher}, \& {Tielens}}]{Pabst2019}
{Pabst}, C., {Higgins}, R., {Goicoechea}, J.~R., {et~al.} 2019, \nat, 565, 618

\bibitem[{{Pabst} {et~al.}(2020){Pabst}, {Goicoechea}, {Teyssier}, {Bern{\'e}}, {Higgins}, {Chambers}, {Kabanovic}, {G{\"u}sten}, {Stutzki}, \& {Tielens}}]{Pabst2020}
{Pabst}, C.~H.~M., {Goicoechea}, J.~R., {Teyssier}, D., {et~al.} 2020, \aap, 639, A2

\bibitem[{{Pound} \& {Wolfire}(2008)}]{Pound2008}
{Pound}, M.~W. \& {Wolfire}, M.~G. 2008, in Astronomical Society of the Pacific Conference Series, Vol. 394, Astronomical Data Analysis Software and Systems XVII, ed. R.~W. {Argyle}, P.~S. {Bunclark}, \& J.~R. {Lewis}, 654

\bibitem[{{Rathjen} {et~al.}(2025){Rathjen}, {Walch}, {Naab}, {N{\"u}rnberger}, {W{\"u}nsch}, {Seifried}, \& {Glover}}]{Rathjen2025}
{Rathjen}, T.-E., {Walch}, S., {Naab}, T., {et~al.} 2025, \mnras [\eprint[arXiv]{2410.00124}]

\bibitem[{{Risacher} {et~al.}(2018){Risacher}, {G{\"u}sten}, {Stutzki}, {H{\"u}bers}, {Aladro}, {Bell}, {Buchbender}, {B{\"u}chel}, {Csengeri}, {Duran}, {Graf}, {Higgins}, {Honingh}, {Jacobs}, {Justen}, {Klein}, {Mertens}, {Okada}, {Parikka}, {P{\"u}tz}, {Reyes}, {Richter}, {Ricken}, {Riquelme}, {Rothbart}, {Schneider}, {Simon}, {Wienold}, {Wiesemeyer}, {Ziebart}, {Fusco}, {Rosner}, \& {Wohler}}]{Risacher2018}
{Risacher}, C., {G{\"u}sten}, R., {Stutzki}, J., {et~al.} 2018, Journal of Astronomical Instrumentation, 7, 1840014

\bibitem[{{Rosolowsky} {et~al.}(2008){Rosolowsky}, {Pineda}, {Kauffmann}, \& {Goodman}}]{Rosolowsky2008}
{Rosolowsky}, E.~W., {Pineda}, J.~E., {Kauffmann}, J., \& {Goodman}, A.~A. 2008, \apj, 679, 1338

\bibitem[{{Rygl} {et~al.}(2012){Rygl}, {Brunthaler}, {Sanna}, {Menten}, {Reid}, {van Langevelde}, {Honma}, {Torstensson}, \& {Fujisawa}}]{Rygl2012}
{Rygl}, K.~L.~J., {Brunthaler}, A., {Sanna}, A., {et~al.} 2012, \aap, 539, A79

\bibitem[{{Röllig, M.} \& {Ossenkopf-Okada, V.}(2022)}]{Roellig2022}
{Röllig, M.} \& {Ossenkopf-Okada, V.} 2022, \aap, 664, A67

\bibitem[{{Schneider} {et~al.}(2023){Schneider}, {Bonne}, {Bontemps}, {Kabanovic}, {Simon}, {Ossenkopf-Okada}, {Buchbender}, {Stutzki}, {Mertens}, {Ricken}, {Csengeri}, \& {Tielens}}]{Schneider2023}
{Schneider}, N., {Bonne}, L., {Bontemps}, S., {et~al.} 2023, Nature Astronomy, 7, 546

\bibitem[{{Schneider} {et~al.}(2016){Schneider}, {Bontemps}, {Motte}, {Blazere}, {Andr{\'e}}, {Anderson}, {Arzoumanian}, {Comer{\'o}n}, {Didelon}, {Di Francesco}, {Duarte-Cabral}, {Guarcello}, {Hennemann}, {Hill}, {K{\"o}nyves}, {Marston}, {Minier}, {Rygl}, {R{\"o}llig}, {Roy}, {Spinoglio}, {Tremblin}, {White}, \& {Wright}}]{Schneider2016a}
{Schneider}, N., {Bontemps}, S., {Motte}, F., {et~al.} 2016, \aap, 591, A40

\bibitem[{{Schneider} {et~al.}(2006){Schneider}, {Bontemps}, {Simon}, {Jakob}, {Motte}, {Miller}, {Kramer}, \& {Stutzki}}]{Schneider2006}
{Schneider}, N., {Bontemps}, S., {Simon}, R., {et~al.} 2006, \aap, 458, 855

\bibitem[{{Schneider} {et~al.}(2010){Schneider}, {Csengeri}, {Bontemps}, {Motte}, {Simon}, {Hennebelle}, {Federrath}, \& {Klessen}}]{Schneider2010}
{Schneider}, N., {Csengeri}, T., {Bontemps}, S., {et~al.} 2010, \aap, 520, A49

\bibitem[{{Schneider} {et~al.}(2020){Schneider}, {Simon}, {Guevara}, {Buchbender}, {Higgins}, {Okada}, {Stutzki}, {G{\"u}sten}, {Anderson}, {Bally}, {Beuther}, {Bonne}, {Bontemps}, {Chambers}, {Csengeri}, {Graf}, {Gusdorf}, {Jacobs}, {Justen}, {Kabanovic}, {Karim}, {Luisi}, {Menten}, {Mertens}, {Mookerjea}, {Ossenkopf-Okada}, {Pabst}, {Pound}, {Richter}, {Reyes}, {Ricken}, {R{\"o}llig}, {Russeil}, {S{\'a}nchez-Monge}, {Sandell}, {Tiwari}, {Wiesemeyer}, {Wolfire}, {Wyrowski}, {Zavagno}, \& {Tielens}}]{Schneider2020}
{Schneider}, N., {Simon}, R., {Guevara}, C., {et~al.} 2020, \pasp, 132, 104301

\bibitem[{{Sofia} {et~al.}(2004){Sofia}, {Lauroesch}, {Meyer}, \& {Cartledge}}]{Sofia2004}
{Sofia}, U.~J., {Lauroesch}, J.~T., {Meyer}, D.~M., \& {Cartledge}, S. I.~B. 2004, \apj, 605, 272

\bibitem[{Sánchez-Monge {et~al.}(2013)Sánchez-Monge, Beltr\'{a}n, ~, Fontani, Brand, Molinari, Testi, \& Burton}]{S_nchez_Monge_2013}
Sánchez-Monge, A., Beltr\'{a}n, M.~T., ~, R., {et~al.} 2013, A\&A, 550, A21

\bibitem[{{Takekoshi} {et~al.}(2019){Takekoshi}, {Fujita}, {Nishimura}, {Taniguchi}, {Yamagishi}, {Matsuo}, {Ohashi}, {Tokuda}, \& {Minamidani}}]{Takekoshi2019}
{Takekoshi}, T., {Fujita}, S., {Nishimura}, A., {et~al.} 2019, \apj, 883, 156

\bibitem[{{Tenorio-Tagle}(1979)}]{Tenorio-Tagle1979}
{Tenorio-Tagle}, G. 1979, \aap, 71, 59

\bibitem[{{Tielens}(2005)}]{Tielens2005}
{Tielens}, A.~G.~G.~M. 2005, {The Physics and Chemistry of the Interstellar Medium}

\bibitem[{{Tiwari} {et~al.}(2021){Tiwari}, {Karim}, {Pound}, {Wolfire}, {Jacob}, {Buchbender}, {Gsten}, {Guevara}, {Higgins}, {Kabanovic}, {Pabst}, {Ricken}, {Schneider}, {Simon}, {Stutzki}, \& {Tielens}}]{Tiwari2021}
{Tiwari}, M., {Karim}, R., {Pound}, M.~W., {et~al.} 2021, \apj, 914, 117

\bibitem[{{Tremblin} {et~al.}(2014){Tremblin}, {Anderson}, {Didelon}, {Raga}, {Minier}, {Ntormousi}, {Pettitt}, {Pinto}, {Samal}, {Schneider}, \& {Zavagno}}]{Tremblin2014b}
{Tremblin}, P., {Anderson}, L.~D., {Didelon}, P., {et~al.} 2014, \aap, 568, A4

\bibitem[{{Walch} {et~al.}(2015){Walch}, {Girichidis}, {Naab}, {Gatto}, {Glover}, {W{\"u}nsch}, {Klessen}, {Clark}, {Peters}, {Derigs}, \& {Baczynski}}]{Walch2015b}
{Walch}, S., {Girichidis}, P., {Naab}, T., {et~al.} 2015, \mnras, 454, 238

\bibitem[{{Walch} {et~al.}(2012){Walch}, {Whitworth}, {Bisbas}, {W{\"u}nsch}, \& {Hubber}}]{Walch2012}
{Walch}, S.~K., {Whitworth}, A.~P., {Bisbas}, T., {W{\"u}nsch}, R., \& {Hubber}, D. 2012, \mnras, 427, 625

\bibitem[{Wang \& Chen(2019)}]{Wang_2019}
Wang, S. \& Chen, X. 2019, The Astrophysical Journal, 877, 116

\bibitem[{Wareing {et~al.}(2018)Wareing, Pittard, Wright, \& Falle}]{Wareing2018}
Wareing, C.~J., Pittard, J.~M., Wright, N.~J., \& Falle, S. A. E.~G. 2018, Monthly Notices of the Royal Astronomical Society, 475, 3598

\bibitem[{{Weaver} {et~al.}(1977){Weaver}, {McCray}, {Castor}, {Shapiro}, \& {Moore}}]{Weaver1977}
{Weaver}, R., {McCray}, R., {Castor}, J., {Shapiro}, P., \& {Moore}, R. 1977, \apj, 218, 377

\bibitem[{{Whitworth} {et~al.}(2022){Whitworth}, {Priestley}, \& {Geen}}]{Whitworth2022}
{Whitworth}, A.~P., {Priestley}, F.~D., \& {Geen}, S.~T. 2022, \mnras, 517, 4940

\bibitem[{{Wright} {et~al.}(2015){Wright}, {Drew}, \& {Mohr-Smith}}]{Wright2015}
{Wright}, N.~J., {Drew}, J.~E., \& {Mohr-Smith}, M. 2015, \mnras, 449, 741

\bibitem[{{W{\"u}nsch} {et~al.}(2021){W{\"u}nsch}, {Walch}, {Dinnbier}, {Seifried}, {Haid}, {Klepitko}, {Whitworth}, \& {Palou{\v{s}}}}]{Wunsch2021}
{W{\"u}nsch}, R., {Walch}, S., {Dinnbier}, F., {et~al.} 2021, \mnras, 505, 3730

\bibitem[{{W{\"u}nsch} {et~al.}(2018){W{\"u}nsch}, {Walch}, {Dinnbier}, \& {Whitworth}}]{Wunsch2018}
{W{\"u}nsch}, R., {Walch}, S., {Dinnbier}, F., \& {Whitworth}, A. 2018, \mnras, 475, 3393

\bibitem[{{Yamagishi} {et~al.}(2018){Yamagishi}, {Nishimura}, {Fujita}, {Takekoshi}, {Matsuo}, {Minamidani}, {Taniguchi}, {Tokuda}, \& {Shimajiri}}]{Yamagishi2018}
{Yamagishi}, M., {Nishimura}, A., {Fujita}, S., {et~al.} 2018, \apjs, 235, 9

\end{thebibliography}
\begin{appendix}

\section{Complementary plots} \label{appendix-plots}

\begin{figure*}[ht]
\begin{center} 
\includegraphics [width=8cm]{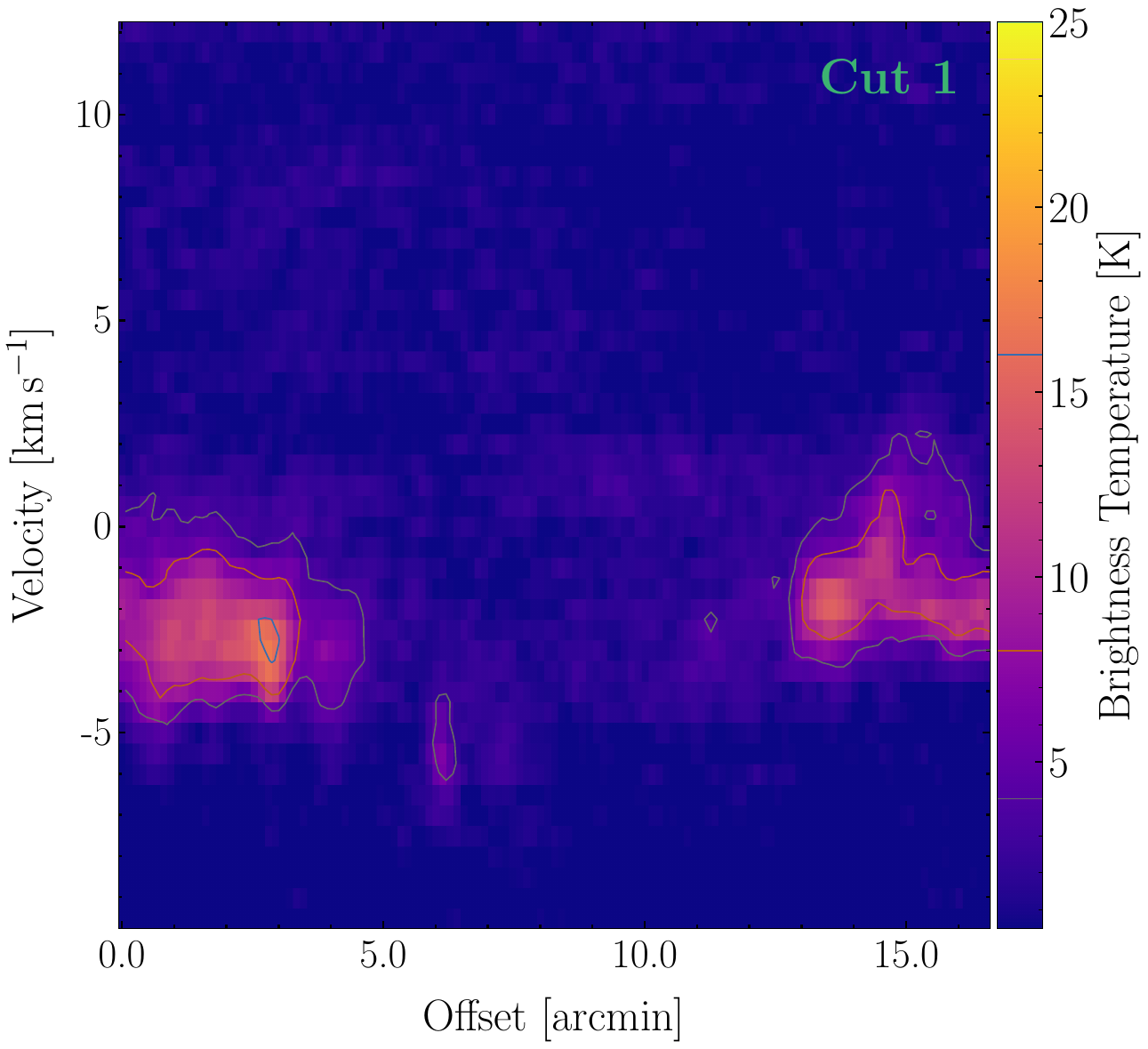}
\includegraphics [width=8cm]{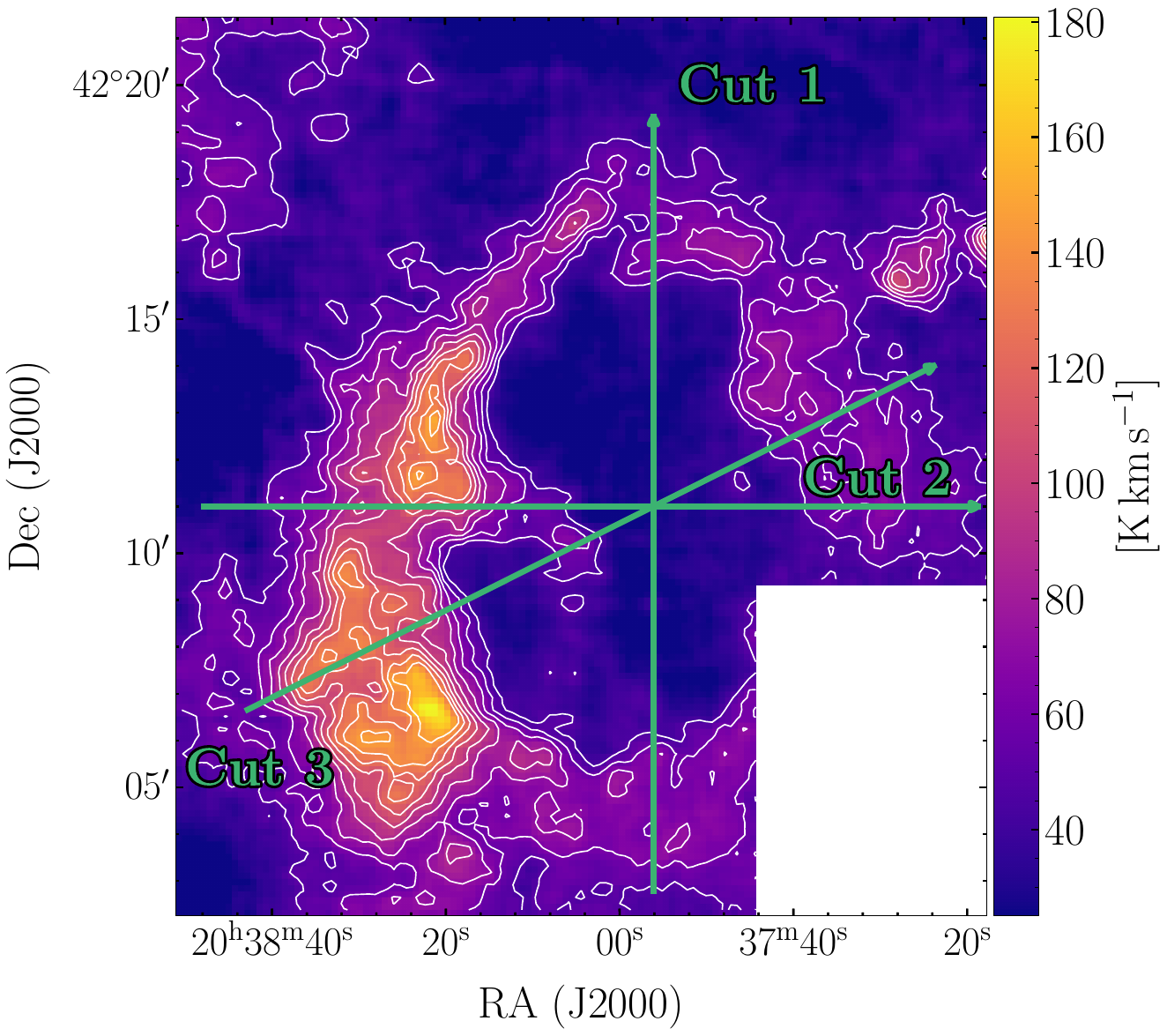}
\includegraphics [width=8cm]{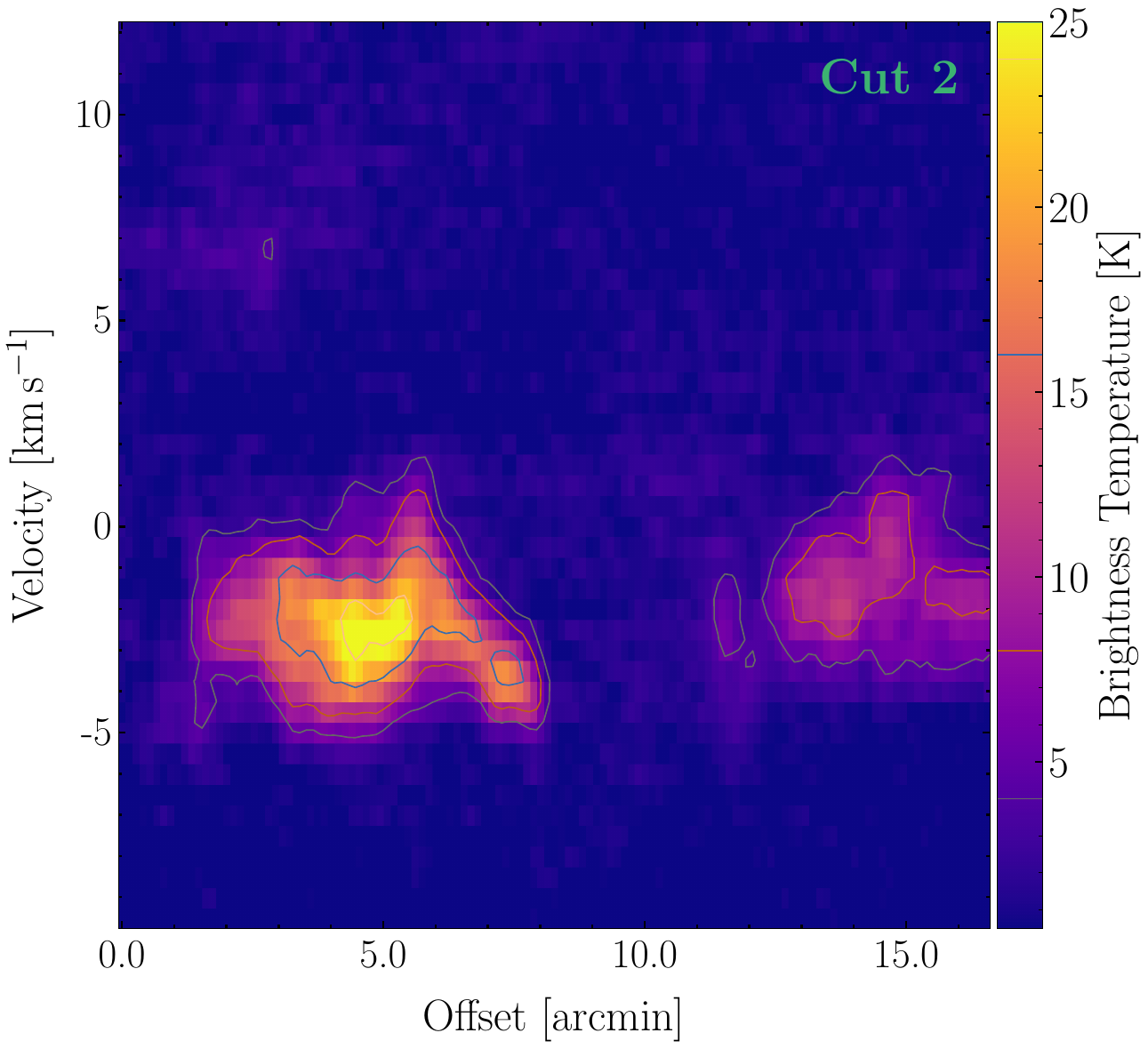}
\includegraphics [width=8cm]{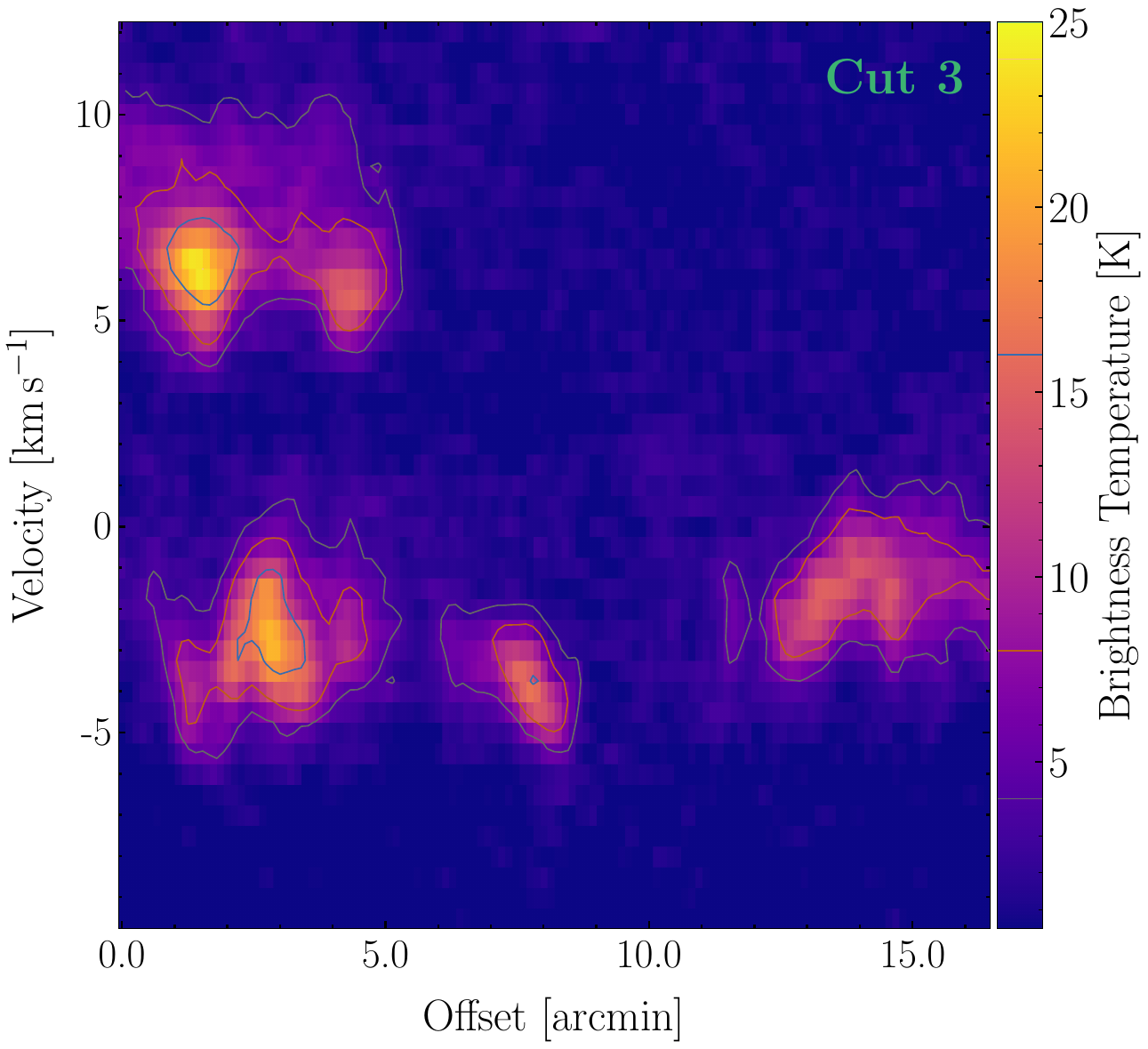}
\caption{Examples of position-velocity cuts in \CII\ of the Diamond Ring from movies that are found \href{https://hera.ph1.uni-koeln.de/~nschneid/Diamond.html}{here}. 
The upper right panel shows the \CII\ line integrated ($-8$ to 20 K km s$^{-1}$) emission in which the contours go from $48$ to $156$ K km s$^{-1}$ in steps of 20 K km s$^{-1}$ ($5\sigma$).
The green lines indicate the cuts, corresponding to the scale in the remaining panels. Movies of the position-velocity cuts are available online.
}
\label{fig:pv-cuts}
\end{center} 
\end{figure*}

\begin{figure}
\begin{centering}
    \includegraphics[width=9cm, angle=0]{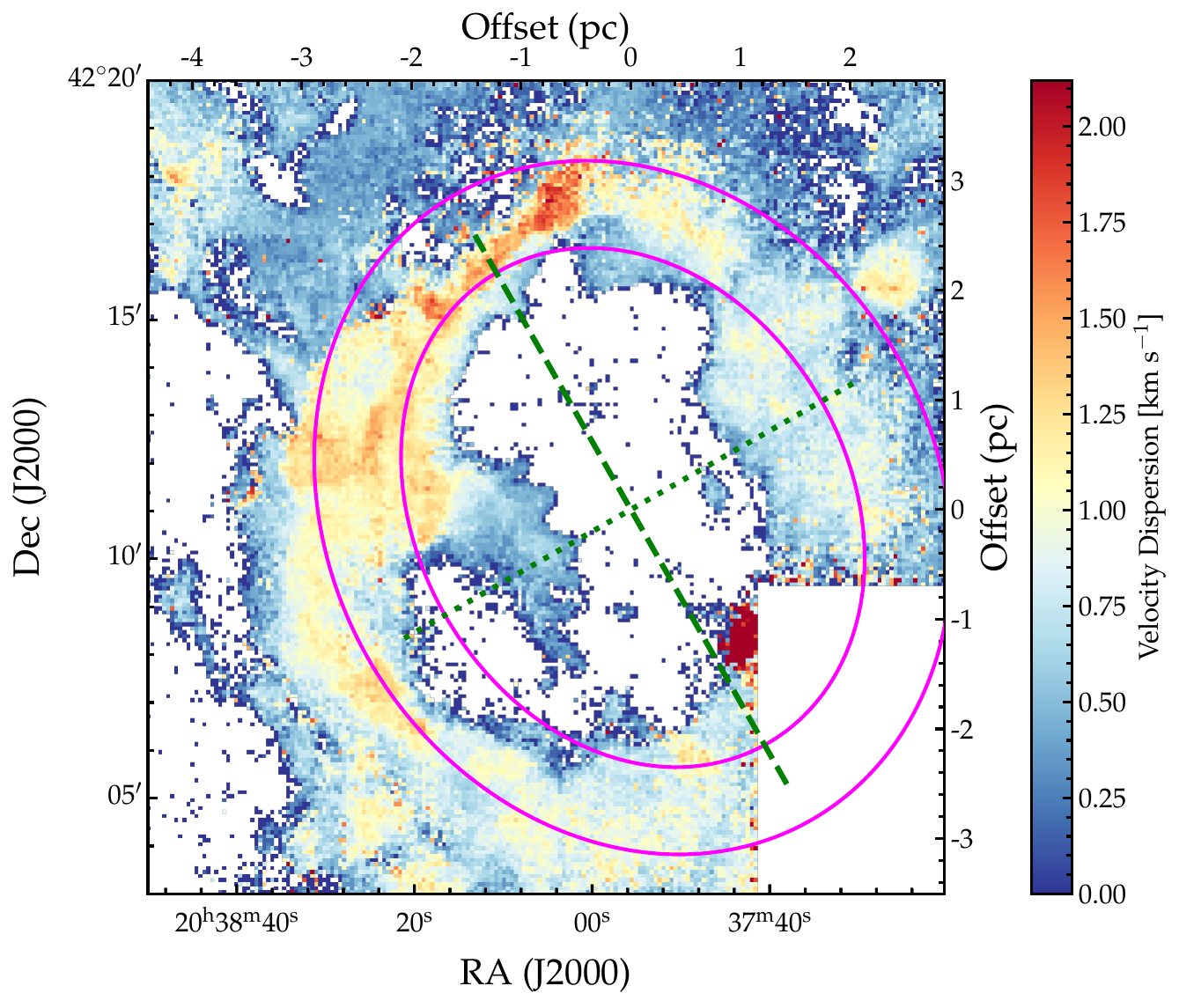}
    \caption{\CII\ moment-2 map (velocity dispersion) of the Diamond Ring, using the same S/N  threshold and fitted ellipse as in the moment-1 velocity map (Figure \ref{fig:ellipse_moment1}).}    \label{fig:ellipse_moment2}
\end{centering}
\end{figure}

Figure~\ref{fig:pv-cuts} displays three position-velocity cuts of the \CII\ emission of the Diamond Ring. The vertical, horizontal, and inclined cuts correspond to the fitted semi minor-axis shown in Fig.~\ref{fig:ellipse_moment1}. None of the cuts show an arc-like emission distribution as it was detected in other sources, such as RCW120 \cite{Luisi2021}, where it is attributed to an expanding \CII\ shell with expansion velocities of $>$10 km\,s$^{-1}$. For the Diamond Ring, the bulk emission shows a linear relation with a slow gradient that increases with redshifted emission, suggesting a LOS velocity difference of $\sim$1.5 km s$^{-1}$, also visible in Fig.~\ref{fig:ellipse_moment1}. This linear relation suggests the expansion of a ring, instead of a three-dimensional shell. In  Appendix~\ref{appendix-ellipse}, we demonstrate that this velocity difference is associated with a radial expansion velocity assuming a tilted ring. Figure \ref{fig:ellipse_moment2} shows the moment-2 map (velocity dispersion) of the \CII\ emission with the same ellipse overlaid as in Fig.~\ref{fig:ellipse_moment1}.

\section{Determination of tilt angle and expansion velocity of the ring} \label{appendix-ellipse}

The relative LOS, $\Delta v_{LOS}$, of the expansion velocity $v_{rad}$ is given by $\Delta v_{LOS}=2 v_{rad}\sin{(\varphi)}$. The factor two arises due to us observing a total velocity difference, so the difference between the approaching and receding parts of the ring and the angle corresponds to the tilt between the plane of the slab and the plane on the sky. If a circular ring is projected onto the sky as an ellipse due to an inclination, this angle is related to the semi-minor axis $b$ and semi-major axis $a$ of an ellipse by $\varphi=\arccos{\left(\frac{b}{a}\right)}$.
An ellipse, accounting for rotation within the plane of sky based on the moment-1 map, was fitted to the crest of the \CII\ emission (assuming a Gaussian emission profile over the ring thickness) at a velocity of $-2.1$ km s$^{-1}$. This velocity corresponds to the maximum emission, as determined by a Gaussian fit to the average spectra over the Diamond Ring and the ring-like appearance is not affected by the SW filament (Fig.~\ref{fig:channels}, left panel). 
A  \texttt{$S/N\geq5$} threshold was applied for fitting. The centre of the ellipse is at RA(2000)=20$^h$37$^m$56$^s$ and Dec(2000)=42$^\circ$11$'$00$''$, approximately 80$''$ separated from star \#227. The major axis of the ellipse is inclined by an angle of 29.6$^\circ$ counter-clockwise with respect to the north direction. This rotation, based on the observed rotation in the moment-1 map, was kept fixed for fitting. The semi-axes have lengths of $b$=2.55~pc and $a$=3.08~pc at a distance of 1.5~kpc. This gives a tilt angle of $\varphi=34^\circ\pm17^\circ$ for the ring rotation. Here, we  assumed an error of 0.4~pc on the axes, which corresponds to roughly half the ring width. The resulting ellipse, overlaid on the \CII\ moment-1 map, is shown in Fig.~\ref{fig:ellipse_moment1}. We determined a LOS velocity difference across the minor axis of the ellipse of $\Delta v_{LOS}$=1.5 km s$^{-1}$ from the associated PV-cut, where we assumed an error of 0.2 km s$^{-1}$. With the equations above, and the associated conservative error estimates, this leads to a low radial expansion velocity of $v_{rad}=1.3\pm0.6$ km~s$^{-1}$. This approach assumes a ring-like geometric structure, whereas, in reality, natural deviations are expected due to expansion into a medium that is never perfectly uniform.

\section{ Ionising source of the associated \HII\ region of the Diamond Ring} \label{appendix-spectra}

\begin{figure}[htbp]
\begin{center}
\includegraphics[width=8.4cm, angle=0]{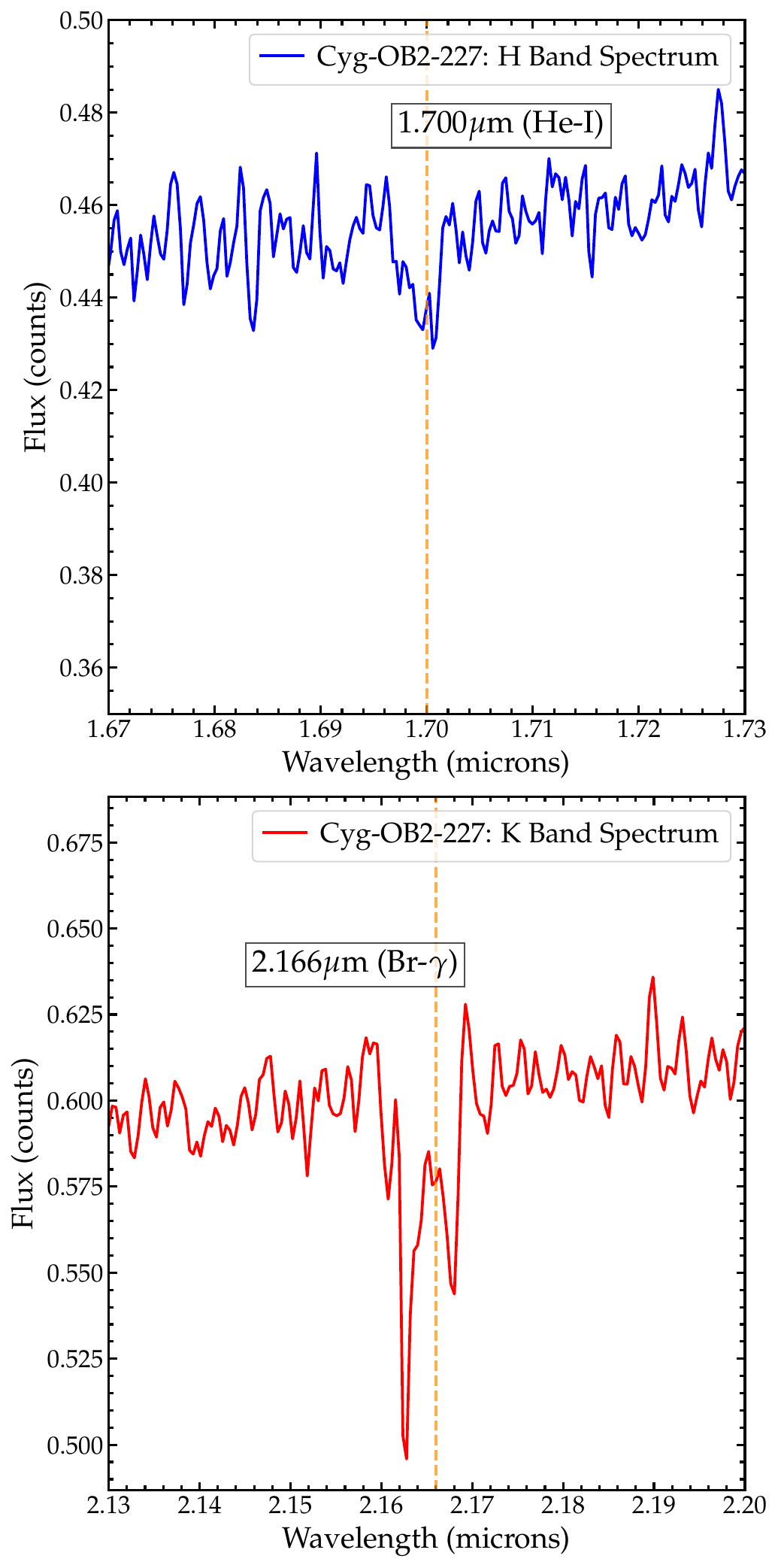}
\vspace{-0.2cm}
\caption{
Regions around the wavelength of the HeI line at 1.700~$\mu$m (top) and of the Brackett gamma line at 2.166~$\mu$m (bottom). Top:  Prominence of the HeI line, together with the absence of the neighbouring HeII line at 1.693 $\mu$m and the absence of the Brackett~11 line at 1.681 $\mu$m, leads us to classify it as a very early B star, presumable a B0.5e spectral type. Bottom:  Brackett gamma line clearly shows a core in emission.
}
\label{fig:spectra_227}
\end{center}
\end{figure}

The location and photometric characteristics of star \#227, at RA(2000)=20$^h$38$^m$05.37$^s$, Dec(2000)=+42$^o$11$'$14.2$''$, make it the obvious candidate to be responsible for the ionisation of the nebula and its subsequent expansion. This is confirmed by its H- and K-band spectrum (Fig~\ref{fig:spectra_227}), which is dominated by the Brackett series of hydrogen in both the H and K bands. In addition, the HeI line at 1.7 
$\mu$m, which quickly decreases in strength with decreasing stellar temperature in main sequence stars, 
is well visible, which suggests a very early B spectral type. On the other hand, we do not detect any HeII features in either the H or the K band, which rules out an O-type. In the K band the HeI feature at 2.112 $\mu$m is also visible, but the most prominent is the Brackett-$\gamma$ line, which shows a prominent emission at its core. We find a spectral type B0.5e as the best match for this star. The photometry is well consistent with this classification and the adopted distance of 1.5 kpc; Gaia DR3 yields $G_{\rm BP} - G_{\rm RP}=3.31$, translating into a visual extinction A$_{\rm v}$ = 8.9 using the relations in \cite{Wang_2019} and an intrinsic colour $(G_{\rm BP} - G_{\rm RP})_{0}=-0.45$. This is similar to the extinction A$_{\rm v}$=9.9 in its direction derived from the {\sl Herschel} column density maps \citep{Schneider2016a}. The absolute magnitude of the star is thus $M_{\rm RP}=-3.7$, or $M_{\rm V}=-4.0$ using the photometric system transformations of \citet{Carrasco2016}, see also  
\href{https://gea.esac.esa.int/archive/documentation/GEDR3/Data_processing/chap_cu5pho/cu5pho_sec\_photSystem/cu5pho\_ssec\_photRelations.html}{GAIA DR3 documentation}. This places the star very near the O9.5-B0 border,
with the caveat that the photometry may be affected by undetected binarity frequently found among massive stars.

\section{Derivation of physical properties of the Diamond Ring} \label{appendix-physicalproperties}

\noindent{{\bf Column density and mass from \CII}} \\ 
For the derivation of the \CII\ column density ($N_{\rm CII}$), we can presume an optically thin $\tau\ll1$ line because the \CII\ line profiles are mostly Gaussian and do not show self-absorption features, as frequently observed in dense and warm PDRs \citep{Kabanovic2022}. We use the equation from \cite{Goldsmith2012}: 
\begin{align}
N_{\rm CII} = \frac{I_{\rm CII}}{3.43\cdot10^{-16}} \cdot \left[1+0.5\cdot e^{\frac{91.25}{T_{kin}}}\left(1+\frac{2.4\cdot10^{-6}}{C_{ul}}\right)\right].
\label{eqn:goldsmith}
\end{align}

We took a kinetic temperature of $T_{kin}$=100 K and a density $n$\footnote{Note: this is the density in the PDR region of the swept-up shell, which is higher than the initial density in the slab of gas (of a few hundred cm$^{-3}$).} after the simulations) of 1000 cm$^{-3}$, based on PDR modeling, using the PDR toolbox and the WK2020 model \citep{Pound2008} of regions within the Diamond Ring. For this modeling, we employed the \CII\ data presented in this study and additional SOFIA observations of the CO 8$\to$7 line and the \OI\ 145 $\mu$m line (to be presented in future publication). 
The de-excitation rate coefficient of atomic hydrogen is given by $R_{ul}=7.6\cdot10^{-19}\left(T_{kin}/100\text{K}\right)^{0.14}\text{cm}^{3}\text{s}^{-1}$, which translates to a collisional de-excitation rate of $C_{ul}=n\times R_{ul}$. 
The calculated column density is then translated to a total mass by 
\begin{align}
M_{\mathrm{gas}}=\frac{N_{\rm CII}}{C/H}m_{\mu} A,
\label{eqn:mass}
\end{align}
with a carbon to hydrogen ratio $C/H=1.6\cdot10^{-4}$ \citep{Sofia2004}, a mean atomic mass 
$m_{\mu}$=$2.2\cdot10^{-27}$ kg, and the area $A$ of the ring. The latter was determined by  performing a Dendrogram \citep{Rosolowsky2008} analysis on the integrated emission map from $-$8 to 3 km s$^{-1}$ and corresponds roughly to the area within the 40 K km s$^{-1}$ level displayed in Fig.~\ref{fig:hii-region}. For the Diamond Ring this yields a value of $N_{\rm CII}=9.7\cdot10^{17}$cm$^{-2}$ and a mass of $\sim$ 1083 M$_{\odot}$ for a mean $I_{\rm CII}$ of 54 K km s$^{-1}$ within the ring area A of $\sim$ 16.9~pc$^2$. \\ 

\noindent{{\bf Lyman continuum flux}} \\ 
From the GLOSTAR cm-observations of the \HII\ region we determine an ionising flux, assuming a single  ionising source, and that all UV photons are reprocessed within the region \citep{S_nchez_Monge_2013}: 
\begin{equation}
\left[\frac{N_{Ly}}{s^{.1}}\right] = 8.9 \times 10^{40}\left[\frac{S_{\nu}}{\unit{Jy}}\right]\left[\frac{\nu}{\unit{GHz}}\right]^{0.1}\left[\frac{T_{e}}{10^4\unit{K}}\right]^{-0.45}\left[\frac{d}{\unit{pc}}\right]^2
.\end{equation}
Here, $S_{\nu}$ is the integrated flux $S_{4.88\unit{GHz}}$= 5.8~Jy of the region inside the Diamond Ring (but excluding cluster Cl~13), $d$ is the distance, and $T_e$ is the electron temperature, assumed to be $8\cdot10^3$~K. For the integrated flux a constant background level of 0.5~Jy has been subtracted, determined as the mean in an emission free region south-west of the Diamond Ring. 
We then obtained a Lyman continuum of $N_{Ly}=1.50\cdot10^{48}$~s$^{-1}$. \\

\noindent{{\bf Age of the \HII\ region}} \\
In order to estimate the age of the \HII\ region associated with the Diamond Ring, we employ results from the simulations of \cite{Tremblin2014b}, which incorporate ram pressure of the surrounding turbulent medium following Larson's law \citep{Larson1981}, in which the \HII\ region expands. They built a grid of 1D simulations to estimate the dynamical age of the \HII\ regions and successfully tested their method on well-known regions for which photometric age estimations were available. In order to derive the ionised gas pressure of the \HII\ region we
first calculate the electron density $\langle n_e \rangle$ following \cite{MartinHernandez2005}:

 \begin{align}
\langle n_e \rangle &= \frac{4.092 \times 10^5 \text{cm}^{-3}}{\sqrt{b(\nu,T_e)}} \left(\frac{S_\nu}{\text{Jy}}\right)^{0.5} \left(\frac{T_e}{10^4\mathrm{K}}\right)^{0.25} \left(\frac{d}{\text{kpc}}\right)^{-0.5} \left(\frac{\theta_D}{\text{"}}\right)^{-1.5}, 
\end{align}
\begin{align}
b(\nu,T_e) &= 1 + 0.3195 \log_{10}\left(\frac{T_e}{10^4\mathrm{K}}\right) - 0.2130 \log_{10}\left(\frac{\nu}{\mathrm{GHz}}\right),
\end{align}
where $\theta_D$ is the angular diameter of the source.  We determined an electron density  $\langle n_e \rangle$ of $\sim86$ cm$^{-3}$ so that the pressure of the ionised gas is $P=2\langle n_e \rangle k_B T_e=1.9\cdot10^{-10}$ dyne cm$^{-2}$. We note that these equations are valid for a thermal pressure-driven \HII\ region without stellar wind. With these values, we obtain an age of the Diamond Ring \HII\ region of  2-2.5 ~Myr, using Fig.~2 of \citet{Tremblin2014b}. We note that if the ionising source is situated within a slab, and the ionised gas is allowed to stream freely into the lower density environment, the method overestimates the age of the region. 
Moreover, the simulations assume a spherical expansion, whereas in this case, the expansion rapidly transitions into the snowplough phase. Here, the gas is primarily driven by the ram pressure of the stellar wind rather than thermal pressure, as the high-pressure gas inside the \HII\ region can freely escape along directions perpendicular to the slab. Consequently, the expansion velocity decreases more rapidly than in a fully confined bubble. As a result, the actual age of the \HII\ region is likely shorter than 2–2.5 Myr. Another timescale commonly discussed in the context of expanding \CII\ bubbles is the dynamical timescale $t_{dyn}$=2.3~Myr, which is just the size-velocity relation and gives an upper-limit, as it assumes a constant expansion speed. 

\end{appendix}
        
\end{document}